\documentclass[twocolumn,times]{aastex701}

\usepackage{soul}
\usepackage{xcolor}
\soulregister\raise7
\soulregister\cite7
\soulregister\ref7
\soulregister\pageref7
\DeclareUnicodeCharacter{02BC}{'}
\usepackage{threeparttable}
\usepackage{txfonts}
\usepackage[utf8]{inputenc}
\newcommand{\dsct}{$\delta$~Scuti }
\begin{document}

\title{Discovery and Characterization of Three New High-Amplitude \dsct Stars from TESS Observations}

\shorttitle{Three New HADS Stars from TESS Observations}
\shortauthors{Lv, C. et al.}

\author[orcid=0000-0001-6354-1646]{Chenglong Lv}
\affiliation{Xinjiang Astronomical Observatory, Chinese Academy of Sciences, Urumqi, Xinjiang 830011, People's Republic of China}
\email[show]{lvchenglong@xao.ac.cn}

\author[orcid=0000-0003-1845-4900]{Ali Esamdin}
\affiliation{Xinjiang Astronomical Observatory, Chinese Academy of Sciences, Urumqi, Xinjiang 830011, People's Republic of China}
\affiliation{School of Astronomy and Space Science, University of Chinese Academy of Sciences, Beijing 100049, People's Republic of China}
\email[show]{aliyi@xao.ac.cn}

\author[orcid=0000-0003-0139-6951]{J.Pascual-Granado}
\affiliation{Instituto de Astrof\'isica de Andaluc\'ia - CSIC, 18008 Granada, Spain}
\email[]{j.pascual@csic.es}

\author[]{Guojie Feng}
\affiliation{Xinjiang Astronomical Observatory, Chinese Academy of Sciences, Urumqi, Xinjiang 830011, People's Republic of China}
\email{fengguojie@xao.ac.cn}

\author[orcid=0000-0002-2463-4943]{Haozhi Wang}
\affiliation{Xinjiang Astronomical Observatory, Chinese Academy of Sciences, Urumqi, Xinjiang 830011, People's Republic of China}
\affiliation{School of Astronomy and Space Science, University of Chinese Academy of Sciences, Beijing 100049, People's Republic of China}
\email{wanghaozhi@xao.ac.cn}

\author{Rivkat Karimov}
\affiliation{Ulugh Beg Astronomical Institute, Uzbekistan Academy of Sciences, Tashkent 100052, Uzbekistan}
\email{rivkat@astrin.uz}

\author{Bolkin Juraev}
\affiliation{Ulugh Beg Astronomical Institute, Uzbekistan Academy of Sciences, Tashkent 100052, Uzbekistan}
\affiliation{Samarkand State University, Samarkand 140104, Uzbekistan}
\email{}

\author{Shuhrat A. Ehgamberdiev}
\affiliation{Ulugh Beg Astronomical Institute, Uzbekistan Academy of Sciences, Tashkent 100052, Uzbekistan}
\affiliation{Samarkand State University, Samarkand 140104, Uzbekistan}
\email{shuhrat@astrin.uz}

\begin{abstract}
We report the discovery and detailed analysis of three new HADS stars, TIC~408074920, TIC~189714989, and TIC~34137913, identified from \textit{TESS} short-cadence observations. Fourier analysis reveals dominant radial modes accompanied by rich harmonic structures in all three stars, confirming their HADS classification.
In addition, TIC~189714989 exhibits clear symmetric side peaks around the radial harmonics, indicative of amplitude and/or phase modulation. Broadband spectral energy distribution (SED) fitting constrained by \textit{Gaia}~DR3 parallaxes provides independent estimates of the stellar effective temperatures and radii. These are compared with results from stellar evolutionary and seismic modelling based on \textsc{MESA} and \textsc{GYRE}. For two targets, the SED-derived and seismic radii are mutually consistent, while TIC~408074920 displays a significant discrepancy, plausibly attributable to the different sensitivities of the two methods and additional systematic effects. These findings highlight the diversity of amplitude variability among classical HADS and emphasize the importance of the long-term monitoring that future missions such as PLATO will allow in order to explore the underlying physical mechanisms of this variability.
\end{abstract}


\section{Introduction}

Over the past two decades, space-based photometric missions have revolutionized stellar astrophysics by providing uninterrupted, high-precision light curves for millions of stars. Early satellites such as \textit{MOST} \citep{Matthews2007}, \textit{CoRoT} \citep{Poretti2009}, and \textit{BRITE} \citep{Weiss2014} pioneered this approach, while \textit{Kepler} delivered continuous four-year monitoring of more than 150,000 stars, establishing unprecedented standards for photometric precision \citep{Borucki2010, Gilliland2010}. More recently, the \textit{Transiting Exoplanet Survey Satellite} (\textit{TESS}; \citealt{Ricker2015}) has extended this legacy through an all-sky survey, generating a vast catalog of variable stars. These efforts have transformed asteroseismology, enabling the detection of subtle oscillations and providing strong constraints on stellar structure and evolution across a broad range of stellar types \citep{Aerts2010, Garcia2019, Daszynska2022, Kurtz2022}.

Within this context, $\delta$~Scuti stars represent a key class of pulsators. These intermediate-mass stars ($1.5$-$2.5~M_\odot$) lie, typically, at the intersection of the classical instability strip and the main sequence, where pressure (p) modes are primarily excited by the opacity ($\kappa$) mechanism operating in the He~\textsc{ii} ionization zone, with an additional significant contribution from turbulent pressure \citep{Antoci2014,Antoci2019}, while gravity (g) modes are thought to be driven by the flux blocking mechanism at the base of the convective envelope, though this excitation process is not yet fully established \citep{Dupret2004,Dupret2005}. Their oscillations provide valuable probes of stellar envelopes and near-core regions, yielding constraints on convective overshooting, internal rotation, and chemical gradients \citep{Poretti2009,Kurtz2014,Lovekin2017,Chen2019}. A particularly informative subgroup are the high-amplitude $\delta$~Scuti (HADS) stars, characterized by large photometric amplitudes ($>0.3$~mag), relatively slow rotation velocity ($v \sin i \lesssim 30$~km/s), and pulsation periods of 1-6 hours \citep{McNamara2000}. Historically, HADS were regarded as nearly pure radial pulsators, dominated by the fundamental and first overtone modes \citep{Petersen1973, Poretti2011}. Their apparent amplitude stability and relatively simple frequency spectra made them long-standing benchmarks for mode identification and for calibrating period-luminosity relations, as well as convenient standard candles for distance determinations \citep{Breger1975,Breger2000,McNamara2000a,Ziaali2019,Mart2022,Jia2025}.

The advent of space-based photometry has significantly altered this picture. Observations from \textit{Kepler} and \textit{TESS} have revealed that many HADS host additional low-amplitude modes beyond the dominant radial pulsations. Sophisticated frequency extraction techniques demonstrate that dozens of weak peaks may be present even in stars once thought to be simple \citep{LaresMartiz2020, Bowman2021, Lv2022,Daszy2024}. This growing evidence suggests that HADS exhibit a richer and more complex oscillation spectrum, raising new questions about nonlinear mode selection and amplitude saturation in intermediate-mass stars. Moreover, long-term, uninterrupted space data have challenged the classical view of amplitude constancy in HADS. While early ground-based campaigns indicated remarkable stability, recent work points to amplitude modulation and even a secular amplitude change in some cases \citep{Poretti2011,Bowman2021,Niu2025}. These findings imply that mode interactions, nonlinear coupling, or evolutionary effects may play a greater role than previously assumed. Understanding whether such behavior is common among HADS, or restricted to rare examples, is essential for placing them within the broader framework of $\delta$~Scuti variability.

Despite these advances, systematic studies of HADS remain scarce. Their intrinsic rarity limits their representation in large surveys, and detailed seismic modeling has so far been attempted for only a handful of stars \citep{Poretti2005,Balona2012,Lovekin2017,Daszynska2020,Bowman2021}. This scarcity hampers our ability to generalize conclusions about their pulsation content, amplitude stability, and evolutionary status. In particular, frequency and period ratios among radial modes offer powerful diagnostics of stellar density and evolutionary stage \citep{Petersen1973, Breger2000}, yet the number of stars with well-determined ratios remains small \citep{Pigulski2006,Balona2016,Netzel2022,Lv2023}. Expanding the well-characterized sample of HADS is therefore crucial for constraining mode excitation mechanisms and for testing stellar evolution theory.

In this work, we report the discovery and analysis of three new HADS stars identified in the \textit{TESS} mission. By analyzing their frequency spectra and amplitude variability, we aim to clarify their role in the $\delta$~Scuti instability strip and to assess their implications for stellar structure and evolution. The paper is organized as follows. Section~\ref{sec:TESS Photometry} details the \textit{TESS} photometry and the methods used for Fourier frequency analysis. Section~\ref{sec:individual_stars} presents the properties of the individual stars, including their light-curve characteristics and pulsation content. Section~\ref{sec:sed} describes the spectral energy distribution fitting and the resulting stellar parameters. Section~\ref{sec:amp} focuses on the analysis of amplitude variability, examining the temporal evolution of pulsation frequencies and amplitudes. Section~\ref{sec:astero} is dedicated to asteroseismic modeling of the targets. Section~\ref{sec:discussion} provides a broader discussion, incorporating constraints from SED consistency checks, placement in the period-luminosity diagram, and evolutionary status from the Hertzsprung-Russell diagram. Finally, Section~\ref{sec:summary} summarizes the main findings and outlines prospects for future work.

\section{TESS Photometry and Fourier Analysis}
\label{sec:TESS Photometry}
\setcounter{footnote}{0}
We adopted the catalog of \citet{Chen2020}, which contains a sample of confirmed $\delta$ Scuti stars identified from \textit{TESS} data, to construct the parent sample for this study. This catalog was cross-matched with \textit{TESS} targets, and the calibrated photometric data processed by the Science Processing Operations Center (SPOC; \citealt{Jenkins2016}) were downloaded using the LIGHTKURVE package \citep{Lightkurve2018}. The data were retrieved from the Mikulski Archive for Space Telescopes (MAST)\footnote{\url{https://archive.stsci.edu/}}. From this cross-matched sample, we searched for HADS and identified three new members of this class. For each target and observing sector, we adopted the PDC-SAP flux time series \citep{Twicken2010}, which corrects common-mode spacecraft systematics and mitigates contaminating flux. Outliers were removed through a $4.5\sigma$ clipping around a running median. Residual long-term trends and ramps were modeled and removed with low-order polynomials fitted to trend-dominated windows, thereby preserving intrinsic variability on timescales relevant to HADS pulsations.

Frequency analyses were performed on the detrended PDC-SAP light curves expressed as relative flux variations. For the 2-min cadence, the Nyquist frequency is $f_\mathrm{N}=360$~c/d, well above the $\delta$~Scuti frequency regime. We searched the range 0–80~c/d, which encompasses the typical domain of HADS pulsations. The formal Rayleigh resolution is $1/\Delta T$, where $\Delta T$ denotes the time span of each combined light curve. Following the criterion suggested by \citet{Loumos1978}, which accounts for the duty cycle and spectral window effects, we adopted a conservative frequency resolution of $f_\mathrm{res}=1.5/\Delta T$ for mode separation. The amplitude spectra were then computed, and an iterative prewhitening analysis was carried out using \textsc{FELIX} \citep{Charpinet2010,Zong2016}. At each iteration the rectified light curve was fitted by a multi-sinusoidal model,
\begin{equation}
x(t)=x_0 + \sum_{i=1}^{N} A_i \sin\!\left[2\pi \left(f_i t + \phi_i\right)\right],
\label{eq:multisine}
\end{equation}

where $x_0$ is the baseline and $(A_i,f_i,\phi_i)$ are the amplitude, frequency, and phase of the $i$-th component. The synthetic signal was subtracted, and the residual spectrum was re-evaluated until no significant peaks remained. Local noise was estimated in a window (2~c/d) excluding regions within $\pm0.05$~c/d of the identified peaks. We considered a frequency significant if its signal-to-noise ratio exceeded S/N $>$ 5.2, which corresponds to the $4\sigma$ false-alarm probability level derived from the individual FAP analyses (see Appendix~A) and is consistent with the criterion adopted by \citet{Baran2015,Baran2021}. Frequency uncertainties were estimated using the analytic expressions of \citet{Montgomery1999}.

\section{Individual Stars}
\label{sec:individual_stars}

\subsection{TIC~408074920}
\label{subsec:tic408074920}

\begin{figure}[htp!]
\begin{center}
  \includegraphics[width=0.5\textwidth]{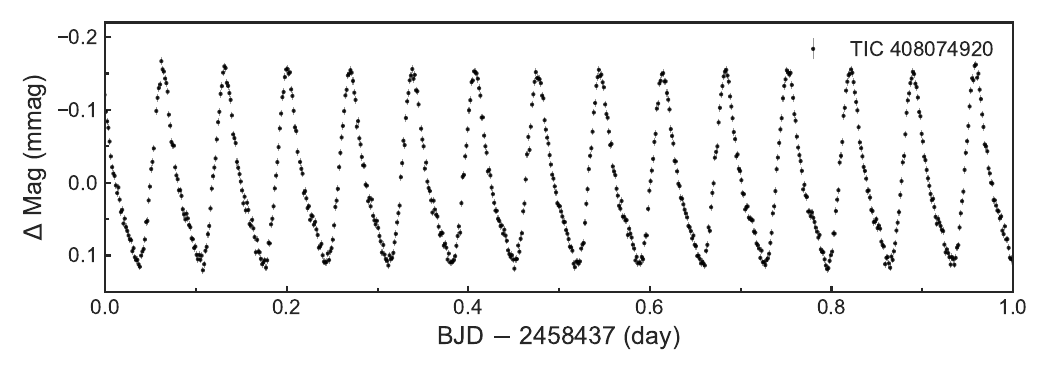}

  \vspace{0.1cm}
  \includegraphics[width=0.5\textwidth]{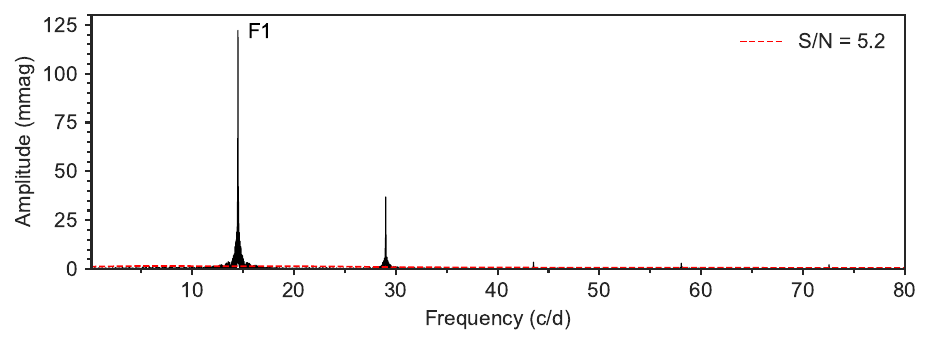}
  \caption{The upper panel displays a one-day segment of the TESS light curve for TIC~408074920. The lower panel presents the corresponding Fourier amplitude spectrum, revealing the dominant frequencies and their harmonics. In the spectrum, the frequency labeled as F1 corresponds to the \textbf{first overtone} radial mode of the star.}
    \label{fig:TIC408074920}
\end{center}
\end{figure}

TIC~408074920 was observed in \textit{TESS} Sector 5 (short cadence; 2 minutes) over a continuous baseline of 27.4 days.  The PDCSAP flux light curve was corrected for instrumental trends and converted to magnitude units. The cleaned light curve exhibits a strongly non-sinusoidal morphology with a peak-to-peak amplitude of approximately 0.3 mag, suggesting significant harmonic content. A one-day segment is illustrated in the upper panel of Figure~\ref{fig:TIC408074920}.

\begin{deluxetable}{cccccc}
\renewcommand\arraystretch{1.2}
\tabletypesize{\small}
\setlength\tabcolsep{18pt}
\tablewidth{\textwidth}
\tablenum{1}
\tablecaption{The Pulsation Mode Frequencies in TESS-SPOC data of TIC 408074920. \label{Tab408074920}}
\tablehead{
\colhead{$f_{i}$} &
\colhead{Frequency (c/d)} &
\colhead{Amplitude (mmag)} &
\colhead{Phase (rad)} &
\colhead{S/N} &
\colhead{Comment}
}
\startdata
$f_1$ & 14.50972(1) & 123.18(8) & 0.7894(2) & 1649.0 & First overtone ($F1$) \\
$f_2$ & 29.01939(4) & 37.01(7)  & 0.7225(6) & 529.8  & $2F1$ \\
$f_3$ & 43.5296(4)  & 3.57(7)   & 0.509(6)  & 51.9   & $3F1$ \\
$f_4$ & 58.0391(5)  & 3.13(7)   & 0.288(7)  & 47.3   & $4F1$ \\
$f_5$ & 72.5486(7)  & 2.08(7)   & 0.13(1)   & 30.8   & $5F1$ \\
$f_6$ & 45.082(3)   & 0.43(7)   & 0.49(5)   & 6.3    & independent mode \\
\enddata
\end{deluxetable}

We performed a standard iterative pre-whitening procedure to extract significant pulsation frequencies. The resulting Fourier amplitude spectrum is displayed in the lower panel of Figure~\ref{fig:TIC408074920}, where the dominant pulsation and its harmonic structure are clearly visible. The extracted frequencies are summarized in Table~\ref{Tab408074920}. The highest-amplitude mode, $f_1 = 14.509721$ c/d (amplitude: 123.2 mmag), is identified as the first overtone radial mode (F1). This identification is supported by the calculated pulsation constant $Q = (1/f)\sqrt{\rho}$, which yields $Q = 0.026 \pm 0.005$ d, where the stellar mean density is derived from \citet{Stassun2019}. According to the commonly adopted classification of radial modes in $\delta$ Scuti stars, fundamental modes typically have $Q \sim 0.033$\,d, while first overtone modes have $Q \sim 0.025$--$0.027$\,d \citep{Stellingwerf1979}. The derived $Q$ value is therefore more consistent with the first overtone radial mode. In the period--luminosity diagram, TIC~408074920 is located between the fundamental and first overtone sequences, without a clear preference for either mode. Harmonics are detected up to $5f_0$. The dominance of a single radial mode and its harmonic series supports the classification of TIC~408074920 as a HADS variable, although the available evidence suggests that the dominant mode is more likely the first overtone. An additional low-amplitude frequency ($f_6$) is detected; however, its limited signal-to-noise ratio and the absence of clear frequency ratios prevent a reliable mode identification at present.

\subsection{TIC~189714989}
\label{subsec:tic189714989}

\begin{figure}[htbp!]
\begin{center}
  \includegraphics[width=0.5\textwidth]{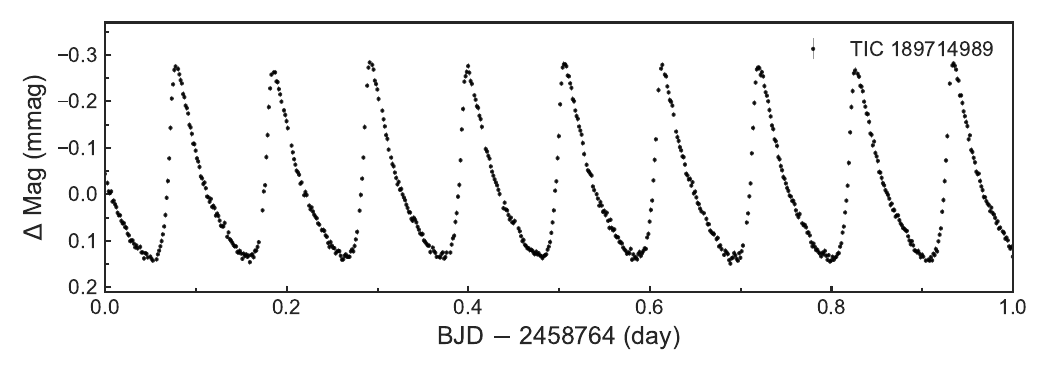}

  \vspace{0.1cm}
  \includegraphics[width=0.5\textwidth]{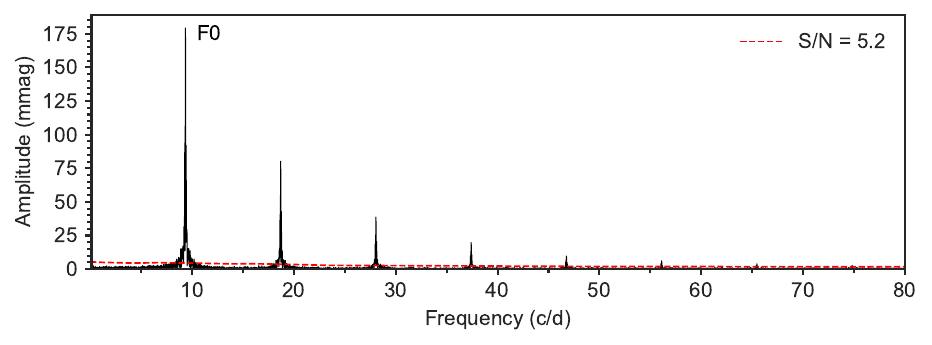}
  \caption{The upper panel displays a one-day segment of the TESS light curve for TIC~189714989. The lower panel presents the corresponding Fourier amplitude spectrum, revealing the dominant frequencies and their harmonics. In the spectrum, the frequency labeled as F0 corresponds to the fundamental radial pulsation mode of the star.}
    \label{fig:TIC189714989}
\end{center}
\end{figure}

TIC~189714989 was observed by \textit{TESS} in Sector 17 (short cadence; 2 minutes) over a baseline of 27 days. The PDCSAP flux was processed using the same reduction procedure as for TIC~408074920, yielding a cleaned light curve that shows large-amplitude ($\sim$0.32 mag) variability with distinctly non-sinusoidal morphology (upper panel of Figure~\ref{fig:TIC189714989}).

\begin{deluxetable}{cccccc}
\renewcommand\arraystretch{1.2}
\tabletypesize{\small}
\setlength\tabcolsep{18pt}
\tablewidth{\textwidth}
\tablenum{2}
\tablecaption{The Pulsation Mode Frequencies in TESS-SPOC data of TIC 189714989. \label{Tab189714989}}
\tablehead{
\colhead{$f_{i}$} &
\colhead{Frequency (c/d)} &
\colhead{Amplitude (mmag)} &
\colhead{Phase (rad)} &
\colhead{S/N} &
\colhead{Comment}
}
\startdata
$f_1$  & 9.35128(2)   & 181.4(1)  & 0.7715(2)  & 1418.2 & Fundamental ($F0$) \\
$f_2$  & 18.70251(3)  & 83.4(1)   & 0.6936(4)  & 798.9  & $2F0$ \\
$f_3$  & 28.05379(5)  & 41.3(1)   & 0.6162(7)  & 443.5  & $3F0$ \\
$f_4$  & 37.40501(9)  & 22.9(1)   & 0.552(1)   & 263.6  & $4F0$ \\
$f_5$  & 46.7563(2)   & 13.4(1)   & 0.477(2)   & 158.4  & $5F0$ \\
$f_6$  & 56.1079(2)   & 9.1(1)    & 0.386(3)   & 111.5  & $6F0$ \\
$f_7$  & 65.4589(3)   & 6.4(1)    & 0.301(4)   & 81.8   & $7F0$ \\
$f_{8}$ & 74.8102(5) & 4.5(1)    & 0.222(6)   & 53.4   & $8F0$ \\
\hline
$f_{9}$  & 9.2306(3)   & 1.0(1)    & 0.636(4)   & 7.8    & $F0 - f_m$ (side peak) \\
$f_{10}$ & 9.4710(6)    & 4.8(1)    & 0.260(9)    & 37.6   & $F0 + f_m$ \\
$f_{11}$ & 18.5820(1)  & 2.3(1)    & 0.508(2)   & 21.5   & $2F0 - f_m$ \\
$f_{12}$ & 18.8224(6)   & 4.5(1)    & 0.197(7)    & 42.8   & $2F0 + f_m$ \\
$f_{13}$ & 27.9342(9)   & 2.5(1)    & 0.449(1)   & 26.5   & $3F0 - f_m$ \\
$f_{14}$ & 28.1733(6)   & 3.7(1)    & 0.131(8)    & 39.6   & $3F0 + f_m$ \\
$f_{15}$ & 37.2840(2)  & 1.3(1)    & 0.469(2)   & 15.4   & $4F0 - f_m$ \\
$f_{16}$ & 37.5244(1)  & 2.1(1)    & 0.079(1)   & 24.6   & $4F0 + f_m$ \\
$f_{17}$ & 46.6336(3)  & 0.67(9)   & 0.447(4)   & 7.9    & $5F0 - f_m$ \\
$f_{18}$ & 46.8756(2)  & 1.1(1)    & 0.025(2)   & 13.2   & $5F0 + f_m$ \\
$f_{19}$ & 55.9854(5)  & 0.44(8)   & 0.296(6)   & 5.4    & $6F0 - f_m$ \\
$f_{20}$ & 56.2267(2)  & 0.88(8)   & 0.888(3)   & 10.7   & $6F0 + f_m$ \\
$f_{21}$ & 65.3435(4)  & 0.44(8)   & 0.051(6)   & 5.6    & $7F0 - f_m$ \\
$f_{22}$ & 65.5783(2)  & 0.86(8)   & 0.795(3)   & 11.1   & $7F0 + f_m$ \\
$f_{23}$ & 74.9304(2)  & 0.95(9)   & 0.651(3)   & 11.2   & $8F0 + f_m$ \\
\hline
$f_{24}$ & 14.1305(2)  & 1.3(1)    & 0.100(3)   & 10.9   & independent mode \\
$f_{25}$ & 23.4837(2)  & 1.1(1)    & 0.984(3)   & 11.4   & $F0 + f_{24}$ (combination)\\
$f_{26}$ & 32.8300(4)  & 0.53(1)   & 0.098(5)   & 5.9    & $2F0 + f_{24}$ \\
\hline
$f_{27}$ & 0.1405(4)    & 8.0(1)    & 0.780(6)    & 56.1   & low-frequency\\
$f_{28}$ & 1.7869(5)   & 0.74(1)   & 0.960(6)   & 5.3    & low-frequency\\
$f_{29}$ & 4.5668(5)   & 0.71(1)   & 0.789(6)   & 5.4    & low-frequency\\
\enddata
\end{deluxetable}

We applied an iterative pre-whitening procedure to extract pulsation frequencies from the TESS light curve of TIC~189714989. The Fourier amplitude spectrum (lower panel of Figure\ref{fig:TIC189714989}) reveals a dense frequency pattern, with 29 frequencies detected above the adopted S/N threshold (Table~\ref{Tab189714989}). The Rayleigh frequency resolution, $1.5/\Delta T \simeq 0.06$~c/d, ensures that all detected peaks are well resolved.

The dominant frequency is $f_1 = 9.35128(2)$~c/d. Using the stellar mean density derived from \citet{Stassun2019}, the corresponding pulsation constant is $Q = 0.034 \pm 0.008$\,d which lies within the range expected for fundamental radial modes. The position of TIC~189714989 on the period-luminosity relation of HADS stars is also consistent with fundamental-mode pulsation. We therefore identify $f_1$ as the fundamental radial mode (F0). A complete sequence of harmonics is present up to 8F0, with observed frequencies closely matching integer multiples of $f_1$ within $10^{-5}$–$10^{-4}$~c/d. This strong harmonic content confirms the star as a HADS pulsator exhibiting strongly nonlinear radial oscillations. In addition to the harmonics, we detect a series of nearly symmetric side peaks around the harmonic frequencies, of the form $n f_1 \pm f_m$, with $f_m\simeq0.1197$~c/d. This modulation frequency generates sidebands around multiple harmonics (e.g., $f_{10} \simeq F_0 + f_{\rm m}$ and $f_{12} \simeq 2F_0 + f_{\rm m}$), indicative of amplitude and/or phase modulation.

The frequency $f_{24}$ = 14.1305(22) c/d is an independent mode detected outside the harmonic and modulation pattern of the dominant radial pulsation. Its period ratio, $P_{24}/P_{0} \approx 0.662$, is markedly lower than the canonical value expected for the first radial overtone in HADS stars ($P_{1}/P_{0} \sim 0.76$--$0.78$) \citep{Stellingwerf1979}, effectively ruling out a simple radial overtone interpretation. Period ratios in the range $P_x/P_0 \sim 0.60$--$0.68$ have indeed been reported in RR~Lyrae stars and are commonly associated with non-radial pulsation modes \citep{BenkHo2023}. However, the frequency structure observed in TIC~189714989 differs fundamentally from the characteristic patterns found in RR~Lyrae variables. In particular, the additional mode $f_{24}$ does not form a distinct frequency family analogous to the $f_{61}$ or $f_{68}$ sequences observed in RR~Lyrae stars, which typically involve multiple related peaks and their combinations with the dominant radial mode. Instead, $f_{24}$ appears as a low-amplitude, isolated frequency that generates only simple linear combinations with the dominant radial mode (e.g. $f_{25} \simeq f_{0} + f_{24}$, $f_{26} \simeq 2f_{0} + f_{24}$). Such behaviour is consistent with a weakly excited low-degree non-radial \mbox{$p$-mode} embedded in a strongly nonlinear radial pulsator. While the excitation of a higher-order radial mode cannot be entirely excluded, the absence of the expected radial-mode period ratios and the observed frequency pattern favour a non-radial interpretation.

Several low-frequency signals are detected at $f_{27}=0.1405$ c/d, $f_{28}=1.7869$ c/d, and $f_{29}=4.5668$ c/d.
Among them, $f_{27}$ is particularly noteworthy.
Given the frequency resolution of the data ($\delta f \approx 0.06$ c/d ), this peak is consistent, within uncertainties, with a modulation frequency.
In Blazhko RR~Lyrae stars, such modulation frequencies are detected either directly as low-frequency peaks or indirectly through symmetric side peaks around the main pulsation modes \citep{Blazhko1907,Yang2019}. In TIC~189714989, symmetric side peaks separated by this frequency are detected around the fundamental mode and all its significant harmonics, which is a characteristic signature of amplitude and/or phase modulation phenomena, similar to the Blazhko effect observed in RR Lyrae stars. Possible physical interpretations of this modulation are discussed in Section~7.

\subsection{TIC~34137913}
\label{subsec:tic34137913}

\begin{figure}[htbp!]
\centering
  \includegraphics[width=0.5\textwidth]{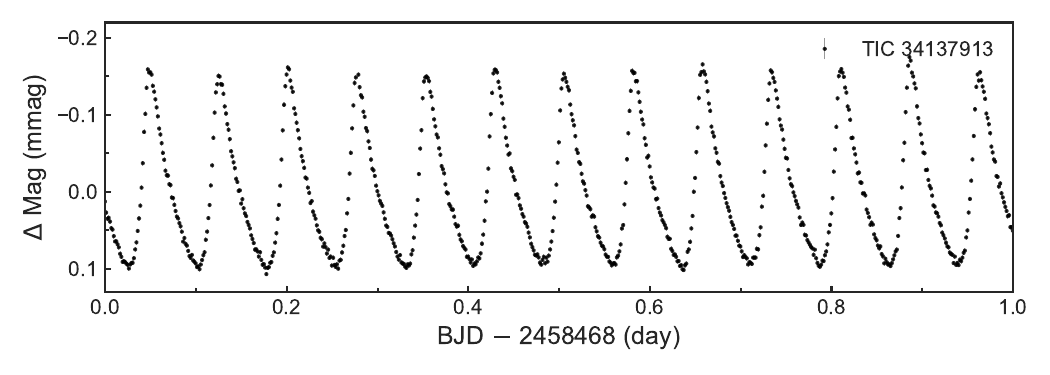}

  \vspace{0.1cm}
  \includegraphics[width=0.5\textwidth]{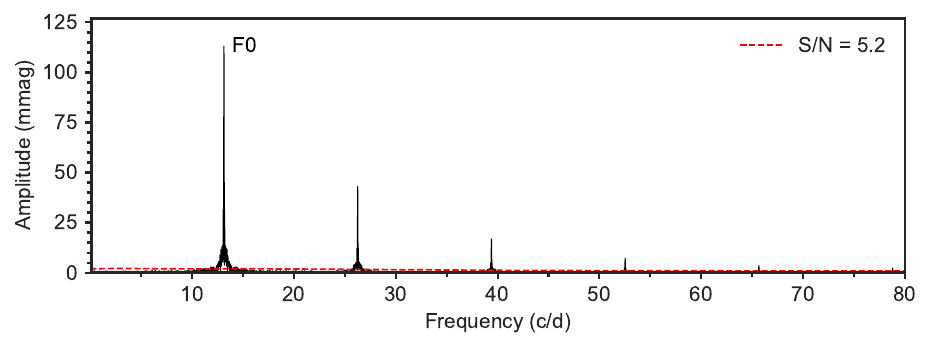}
  \caption{The upper panel displays a one-day segment of the TESS light curve for TIC~34137913. The lower panel presents the corresponding Fourier amplitude spectrum, revealing the dominant frequencies and their harmonics. In the spectrum, the frequency labeled as F0 corresponds to the fundamental radial pulsation mode of the star.}
  \label{fig:TIC34137913}
\end{figure}

TIC~34137913 was observed in \textit{TESS} Sector 6, providing a high-quality PDCSAP light curve. The processed light curve exhibits a large peak-to-peak variability exceeding 0.25 mag (upper panel of Figure~\ref{fig:TIC34137913}).

We applied an iterative pre-whitening procedure to extract significant frequencies. The resulting Fourier amplitude spectrum (lower panel of Figure~\ref{fig:TIC34137913}) shows 10 prominent peaks above the S/N threshold (Table~\ref{Tab34137913}). Given the continuous time base of the observations, the Rayleigh frequency resolution is $1.5/\Delta T \simeq 0.05$ c/d, ensuring that all detected frequencies are well resolved. The dominant frequency, $f_1 = 13.131429$ c/d, is identified as the fundamental radial mode (F0), supported by its pulsation constant($0.042 \pm 0.006$ d) and its position on the period-luminosity relation of HADS stars, with harmonics clearly detected up to 6F0. An additional independent frequency, $f_7 = 17.6174$ c/d, is also present. To investigate its nature, we computed the period ratio $P_7 / P_0 \sim 0.746$. The observed ratio is slightly below the canonical range for the first overtone, suggesting that $f_7$ is unlikely a perfect first overtone radial mode and may correspond to a low-degree nonradial p-mode or a slightly higher-order radial mode.

Several combination frequencies involving $f_1$ and $f_7$ are detected, including $f_8 \simeq f_1 + f_7$, $f_9 \simeq 2f_7 - f_1$, and $f_{10} \simeq 2f_1 + f_7$, supporting the astrophysical origin of $f_7$ and indicating nonlinear mode interactions. The coexistence of a dominant radial mode and an additional independent mode, which may be nonradial or a higher-order radial mode, is observed in TIC~34137913.

\begin{deluxetable}{cccccc}
\renewcommand\arraystretch{1.2}
\tabletypesize{\small}
\setlength\tabcolsep{18pt}
\tablewidth{\textwidth}
\tablenum{3}
\tablecaption{The Pulsation Mode Frequencies in TESS-SPOC data of TIC 34137913. \label{Tab34137913}}
\tablehead{
\colhead{$f_{i}$} &
\colhead{Frequency (c/d)} &
\colhead{Amplitude (mmag)} &
\colhead{Phase (rad)} &
\colhead{S/N} &
\colhead{Comment}
}
\startdata
$f_1$  & 13.13143(3)  & 114.2(1)  & 0.6543(3)  & 977.1 & Fundamental ($F0$) \\
$f_2$  & 26.26290(4)  & 44.16(1)  & 0.4601(5)  & 597.5 & $2F0$ \\
$f_3$  & 39.39414(9)  & 18.16(1)  & 0.256(1)   & 297.0 & $3F0$ \\
$f_4$  & 52.52587(2)  & 8.58(1)   & 0.052(2)   & 155.1 & $4F0$ \\
$f_5$  & 65.65735(3)  & 4.48(1)   & 0.842(4)   & 82.8  & $5F0$ \\
$f_6$  & 78.78826(6)  & 2.46(1)   & 0.637(7)   & 46.4  & $6F0$ \\
\hline
$f_7$  & 17.617(2)    & 1.0(1)    & 0.57(3)    & 10.4  & $\mathrm{independent\ mode}$ \\
$f_8$  & 30.749(3)    & 0.62(7)   & 0.35(4)    & 9.0   & $F0 + f_7$ (combination) \\
$f_9$  & 22.189(4)    & 0.50(8)   & 0.00(5)    & 5.9   & $2f_7 - F0$  \\
$f_{10}$ & 43.878(4)  & 0.37(6)   & 0.17(5)    & 6.4   & $2F0 + f_7$  \\
\enddata
\end{deluxetable}

\section{Spectral Energy Distribution Fitting}
\label{sec:sed}

To determine the fundamental stellar properties of TIC~408074920,
TIC~189714989, and TIC~34137913, we carried out a broad-band SED analysis. This approach provides an independent and robust means of constraining stellar parameters by exploiting multi-wavelength photometric measurements that sample the stellar flux from the ultraviolet through the infrared. The fitting procedure was implemented with the \texttt{Python}-based tool \texttt{ASTROARIADNE}\footnote{\url{https://github.com/jvines/astroARIADNE}} \citep{Vines2022}, which combines a Nested Sampling framework with a suite of atmospheric model grids to deliver statistically rigorous posterior distributions of the stellar parameters. In \texttt{ASTROARIADNE}, parameter estimation is performed using the \texttt{dynesty} nested sampling algorithm \citep{Speagle2020}, which efficiently explores the parameter space and generates statistically meaningful posterior distributions for all fitted parameters \citep{Vines2022}.

\begin{deluxetable*}{l c c c}
\renewcommand\arraystretch{1.2}
\tabletypesize{\small}
\tablewidth{\textwidth}
\tablenum{4}
\tablecaption{Photometric magnitudes of TIC~408074920, TIC~189714989, and TIC~34137913. \label{Tab:sed}}
\tablehead{
\colhead{Bands} &
\colhead{TIC 408074920} &
\colhead{TIC 189714989} &
\colhead{TIC 34137913} \\[-0.5ex]
\colhead{} &
\colhead{(mag)} &
\colhead{(mag)} &
\colhead{(mag)}
}
\startdata
$BP$  & 13.01 $\pm$ 0.02 & 12.94 $\pm$ 0.05 & 12.51 $\pm$ 0.03 \\
$G$   & 12.81 $\pm$ 0.01 & 12.76 $\pm$ 0.01 & 12.25 $\pm$ 0.01 \\
$RP$  & 12.49 $\pm$ 0.01 & 12.52 $\pm$ 0.03 & 11.82 $\pm$ 0.02 \\
$TESS$    & 12.54 $\pm$ 0.01 & 12.57 $\pm$ 0.02 & 11.92 $\pm$ 0.02 \\
$y_{ps}$  & 12.87 $\pm$ 0.01 & -         & 12.22 $\pm$ 0.01 \\
$J$       & 12.10 $\pm$ 0.02 & 12.12 $\pm$ 0.02 & 11.29 $\pm$ 0.02 \\
$H$       & 11.98 $\pm$ 0.03 & 12.06 $\pm$ 0.02 & 11.10 $\pm$ 0.02 \\
$Ks$      & 11.89 $\pm$ 0.02 & 12.01 $\pm$ 0.02 & 11.06 $\pm$ 0.02 \\
W1  & 11.83 $\pm$ 0.02 & 12.03 $\pm$ 0.02 & 10.94 $\pm$ 0.02 \\
W2  & 11.84 $\pm$ 0.02 & 12.06 $\pm$ 0.02 & 10.94 $\pm$ 0.02 \\
\enddata
\tablecomments{The table lists the multi-band photometry of three targets. Missing data are denoted as '-'.}
\end{deluxetable*}

The construction of the observed SEDs began with the collection of broadband photometry from a wide range of archival surveys, automatically queried through the \texttt{astroquery} package. The adopted data sets encompass ultraviolet to mid-infrared passbands, specifically: \textit{GALEX} \citep{Martin2005, Bianchi2011}, Sloan Digital Sky Survey \citep[SDSS;][]{Abazajian2009}, the AAVSO Photometric All-Sky Survey \citep[APASS;][]{Henden2014}, \citep[Pan-STARRS;][]{Chambers2016}, the \textit{TESS} mission \citep{Ricker2015}, 2MASS \citep{Skrutskie2006}, \textit{Kepler} \citep{Borucki2010}, NGTS \citep{Wheatley2018}, and the \textit{ALLWISE} data release \citep{Wright2010}. The compiled magnitudes across these surveys, which provide the foundation for the SED fitting, are listed in Table~\ref{Tab:sed} for the three stars under study.

In order to break the degeneracies between temperature, radius, and extinction that are intrinsic to SED modeling, we incorporated the trigonometric parallaxes from \textit{Gaia} DR3 as informative priors on the stellar distances. If a parallax measurement is available, \texttt{ASTROARIADNE} uses it to constrain the distance by applying a Gaussian prior centered on the inverse of the measured parallax, with the standard deviation equal to the parallax uncertainty. This prior ensures a physically consistent luminosity–radius relation and helps to break the degeneracy among $T_{\mathrm{eff}}$, $R_\star$, and $A_V$. The likelihood function used by \texttt{ASTROARIADNE} is defined as a $\chi^2$ comparison between the observed and model fluxes across all passbands, weighted by the photometric errors. This ensures that the resulting posterior distributions are statistically consistent with both photometric and astrometric constraints.

For the model fitting, \texttt{ASTROARIADNE} employs five families of stellar atmosphere models: \texttt{phoenix v2} \citep{Husser2013}, BT-Settl \citep{Allard2012}, BT-Cond \citep{Allard2012}, BT-NextGen \citep{Hauschildt1999,Allard2012}. These model grids cover $T_{\mathrm{eff}}$ from 2300~K to 12000~K and metallicities in the range $-2.5 < \mathrm{[Fe/H]} < +0.5$, and have been validated for A-F type stars \citep{Murphy2022}. By simultaneously sampling from multiple atmospheric grids, the code mitigates the biases that can arise when relying on a single model set, thereby improving the reliability of the derived parameters. For each star, the posterior distributions yield estimates of the effective temperature ($T_{\mathrm{eff}}$), surface gravity ($\log g$), metallicity ([Fe/H]), stellar radius, extinction in the $V$ band ($A_V$), and distance. The derived parameters for the three targets are summarized in Table~\ref{Tab:SEDresults}.

\begin{deluxetable}{cccccc}
\renewcommand\arraystretch{1.2}
\tabletypesize{\small}
\setlength\tabcolsep{22pt}
\tablewidth{\textwidth}
\tablenum{5}
\tablecaption{Fundamental parameters derived from SED fitting for the three HADS stars.\label{Tab:SEDresults}}
\tablehead{
\colhead{TIC ID} &
\colhead{$T_{\mathrm{eff}}$ (K)} &
\colhead{$\log g$} &
\colhead{[Fe/H]} &
\colhead{$R_\star (R_\odot)$} &
\colhead{$A_V$ (mag)}
}
\startdata
408074920 & $6994^{+30}_{-36}$   & $3.76^{+0.08}_{-0.07}$ & $-0.21^{+0.09}_{-0.09}$ & $2.64^{+0.06}_{-0.07}$ & $0.27^{+0.01}_{-0.01}$ \\
189714989 & $7167^{+29}_{-30}$   & $4.10^{+0.03}_{-0.03}$ & $-0.15^{+0.08}_{-0.08}$ & $2.65^{+0.07}_{-0.07}$ & $0.08^{+0.01}_{-0.01}$ \\
34137913  & $7406^{+158}_{-151}$ & $4.02^{+0.26}_{-0.28}$ & $-0.15^{+0.13}_{-0.13}$ & $2.06^{+0.07}_{-0.08}$ & $0.97^{+0.11}_{-0.11}$ \\
\enddata
\end{deluxetable}

The derived parameters are generally consistent with the stars being main-sequence or mildly evolved A-F type stars. The effective temperatures and radii fall within the ranges expected for $\delta$~Scuti stars, and the residuals observed in the SED fits indicate a satisfactory agreement between the synthetic models and the multi-band photometry. Within the limitations inherent to SED-based analyses, these results provide a useful set of global stellar parameters that serve as an independent reference for interpreting the pulsational characteristics of these targets.

\begin{figure*}[htbp!]
\centering
\includegraphics[width=0.45\textwidth]{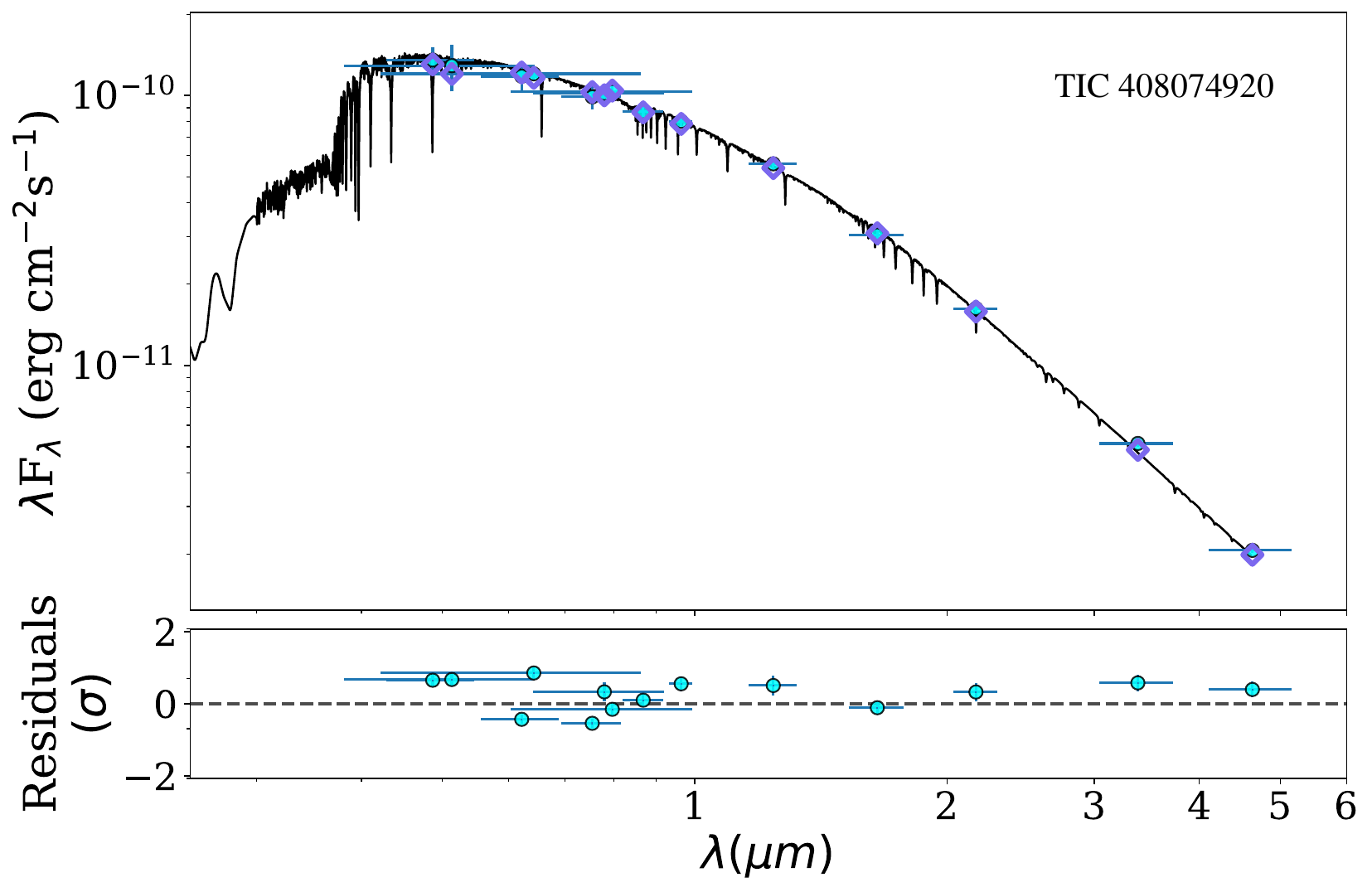}
\includegraphics[width=0.4\textwidth]{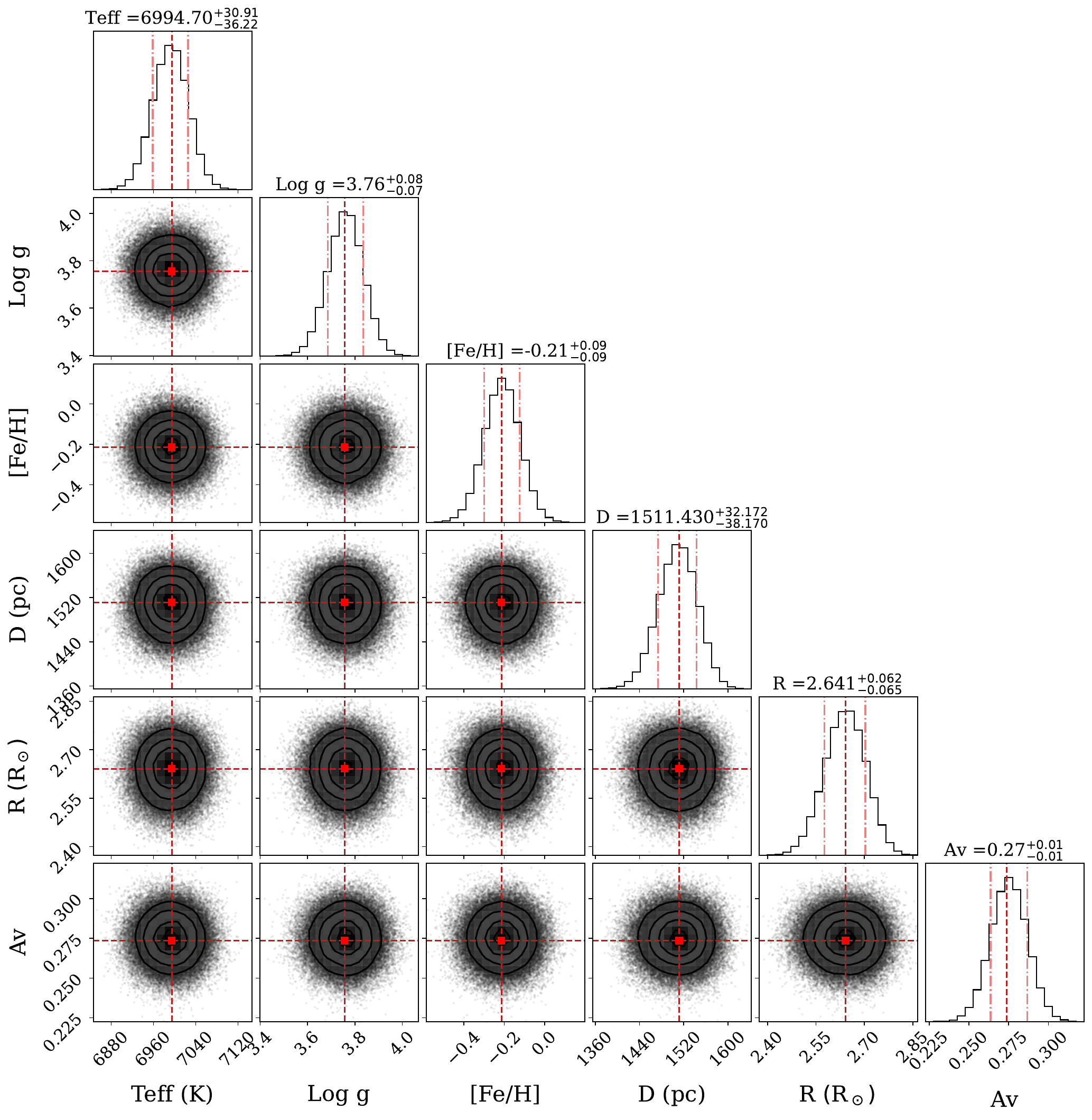}\\[3pt]
\includegraphics[width=0.45\textwidth]{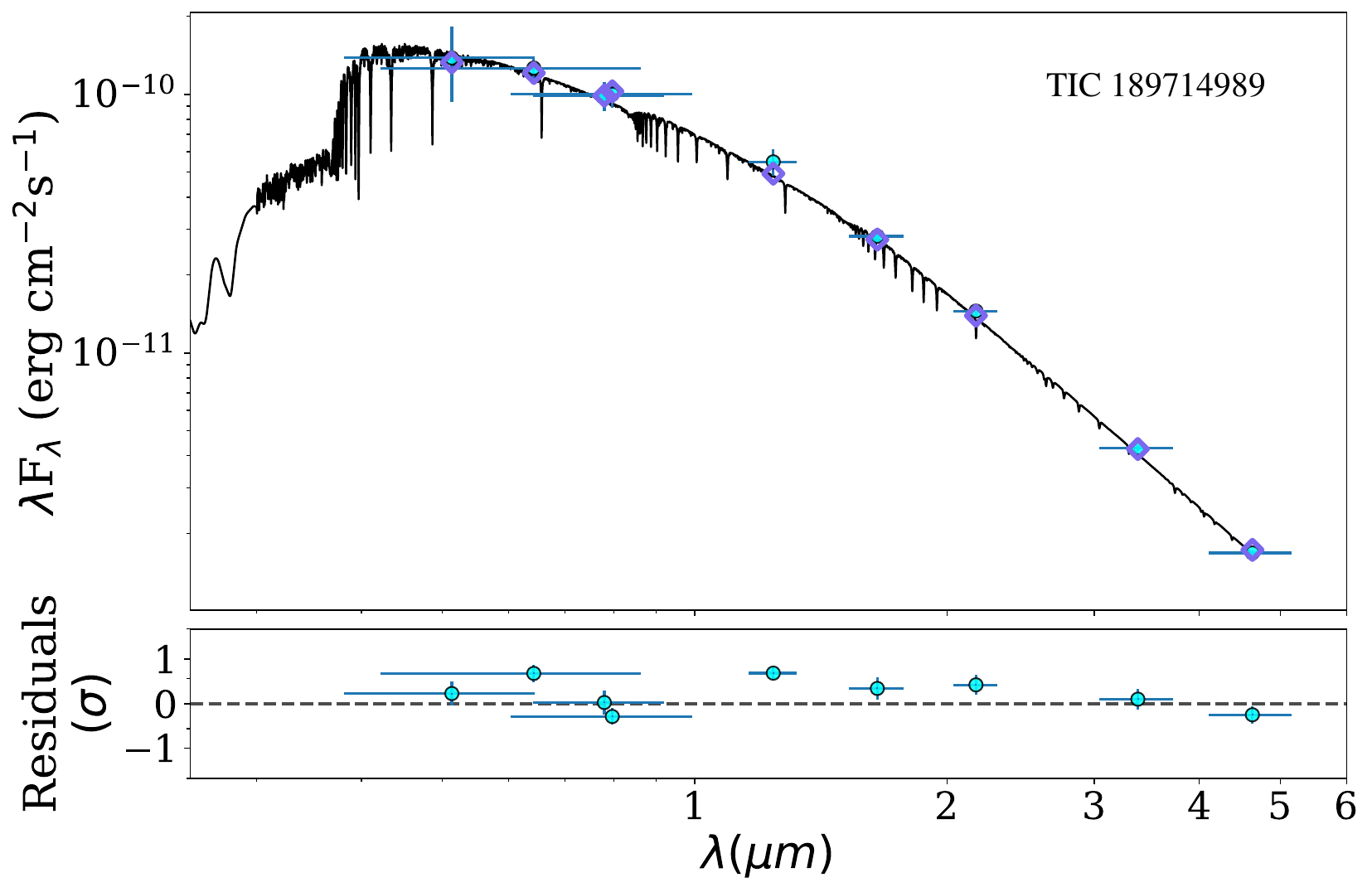}
\includegraphics[width=0.4\textwidth]{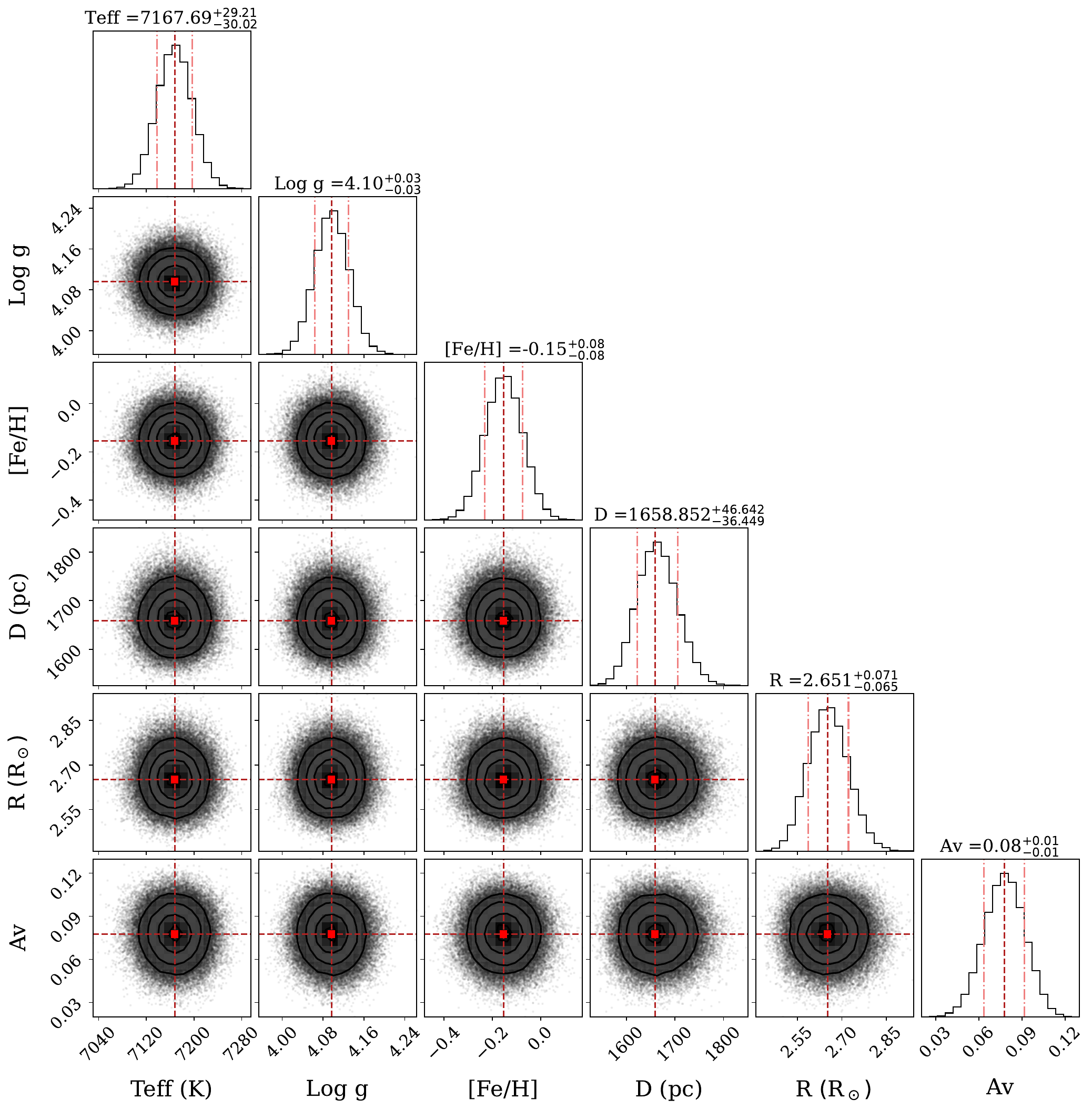}\\[3pt]
\includegraphics[width=0.45\textwidth]{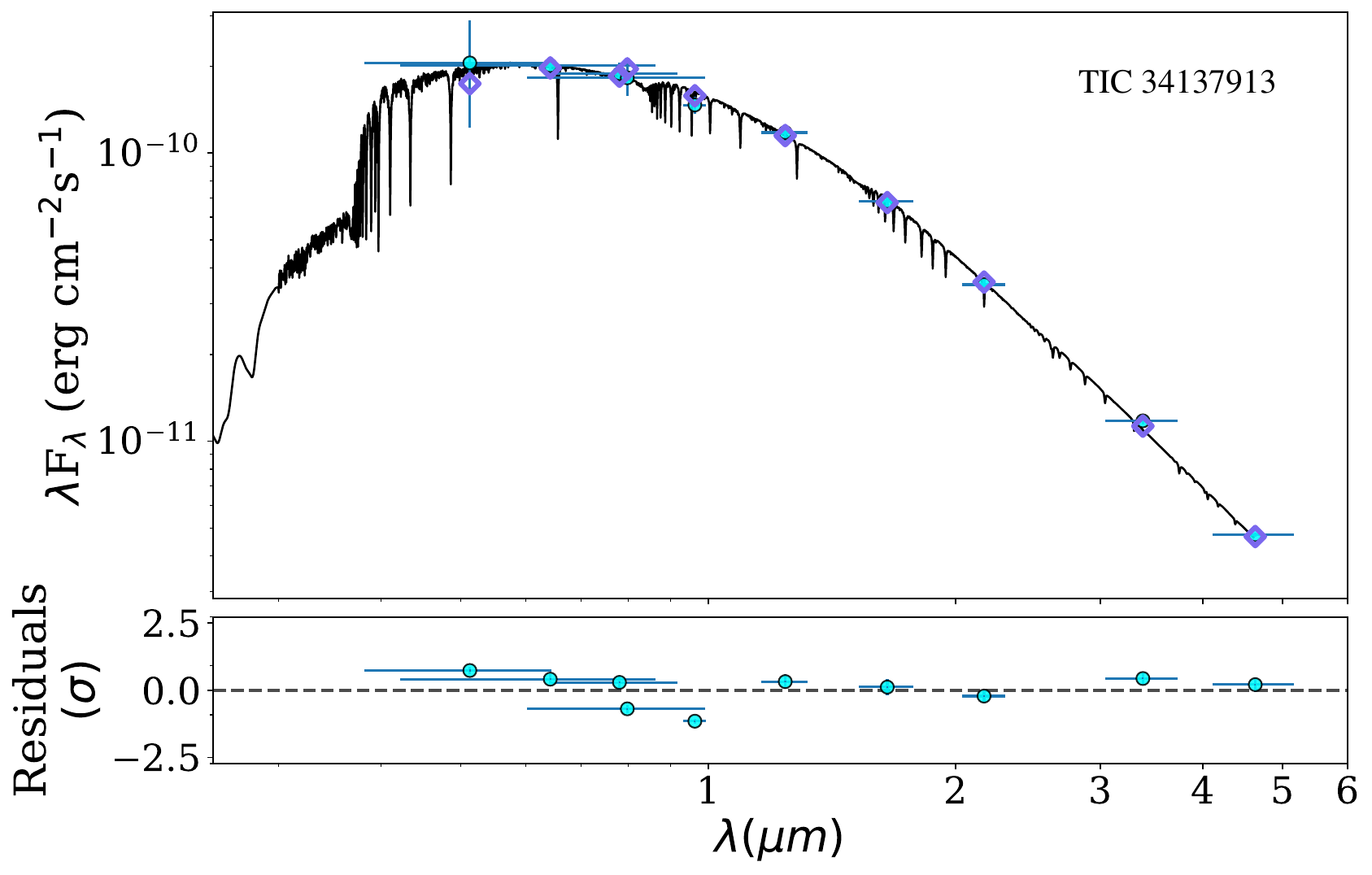}
\includegraphics[width=0.4\textwidth]{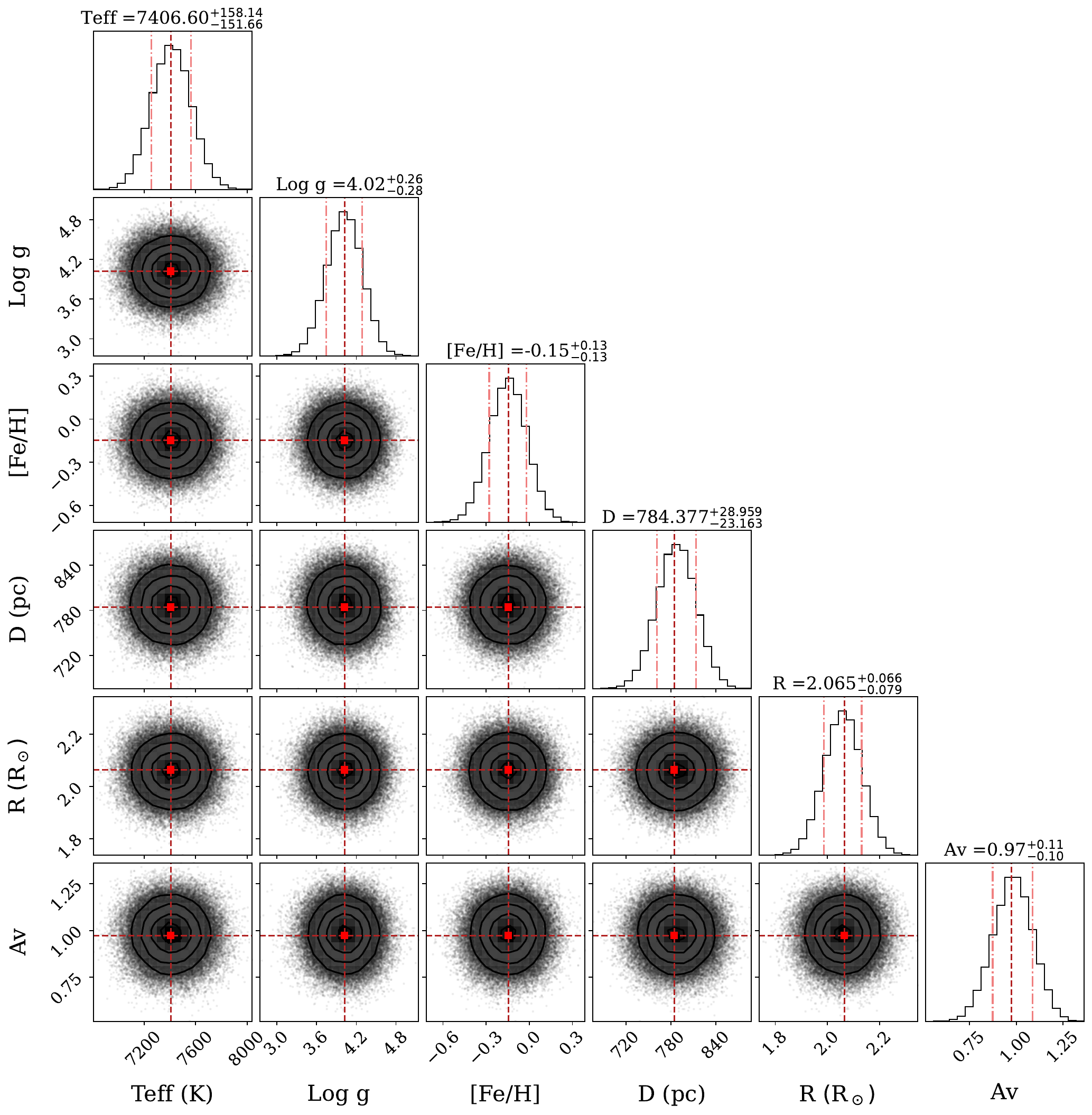}
\caption{Results of SED fitting for the three HADS stars.
\textit{Left panels:} Observed fluxes (cyan points) and best-fit model fluxes (black curves).
\textit{Right panels:} Corner plots showing posterior distributions of the fitted parameters ($T_{\mathrm{eff}}$, $\log g$, [Fe/H], distance, radius, and $A_V$).}
\label{fig:SED_corner_all}
\end{figure*}

Figures~\ref{fig:SED_corner_all} present the SED fitting results and the corresponding corner plots for all targets. Each figure shows the observed photometric measurements compared with the model SEDs, providing a clear visual assessment of the fitting quality. The corner plots illustrate the posterior probability distributions of the fitted parameters and highlight the degree of correlation among them, demonstrating the statistical consistency of the derived stellar properties. The relatively small uncertainties seen in the corner plots mainly reflect the strong constraints provided by the broad-band SED shape and the precise Gaia distances, particularly for $T_{\mathrm{eff}}$ and radius. However, parameters such as $\log g$, which are only weakly constrained by SED fitting, are more sensitive to model assumptions and adopted priors. Therefore, their formal uncertainties should be interpreted as statistical precision and may underestimate the true uncertainties due to potential systematic effects.

\section{Amplitude Variability Analysis}
\label{sec:amp}

\begin{figure*}[htbp!]
\centering
\setlength{\tabcolsep}{2pt}
\begin{tabular}{ccc}
\includegraphics[width=0.33\textwidth]{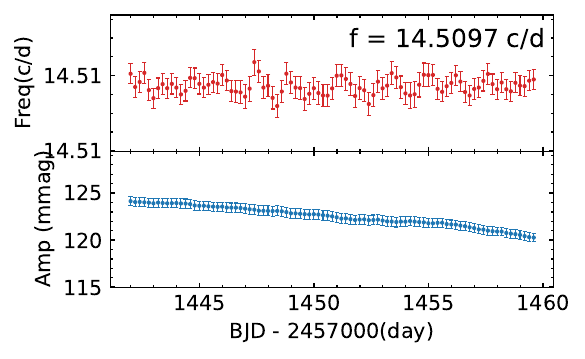} &
\includegraphics[width=0.33\textwidth]{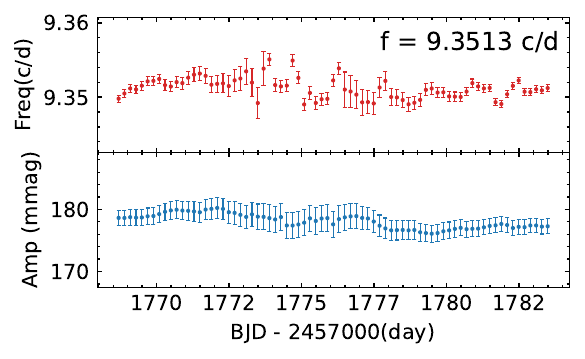} &
\includegraphics[width=0.33\textwidth]{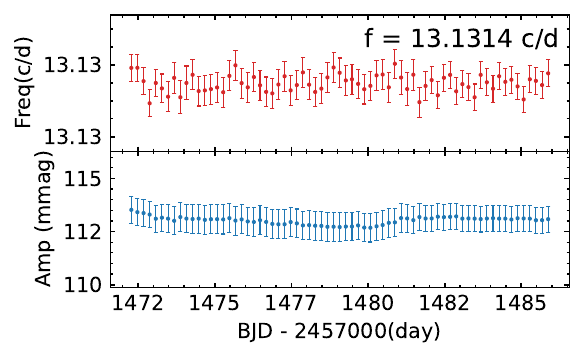} \\[-2pt]
\includegraphics[width=0.33\textwidth]{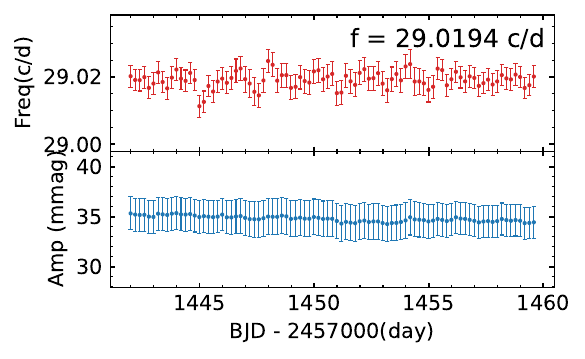} &
\includegraphics[width=0.33\textwidth]{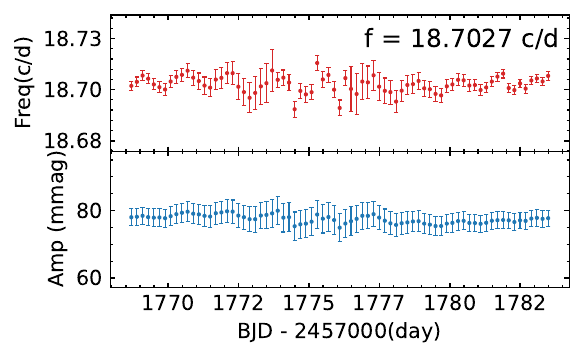} &
\includegraphics[width=0.33\textwidth]{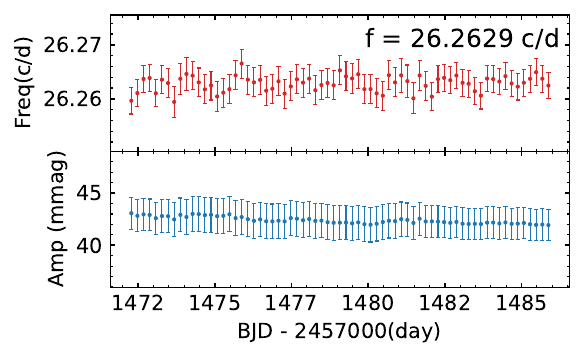} \\[-2pt]
\includegraphics[width=0.33\textwidth]{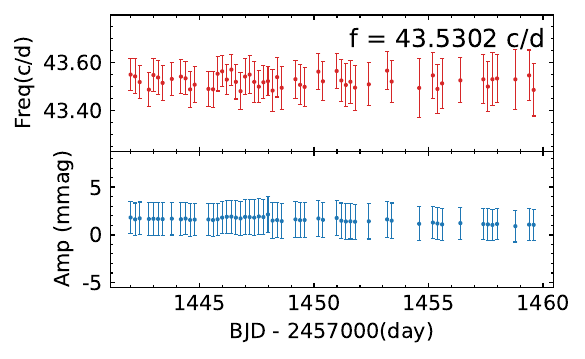} &
\includegraphics[width=0.33\textwidth]{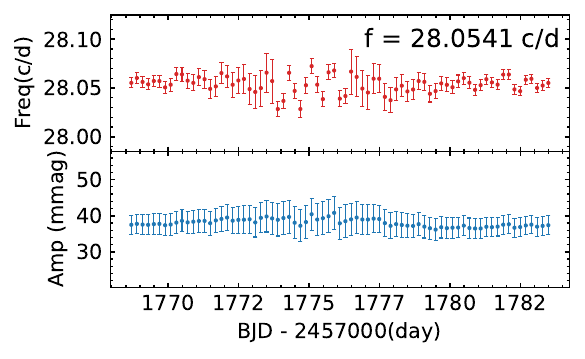} &
\includegraphics[width=0.33\textwidth]{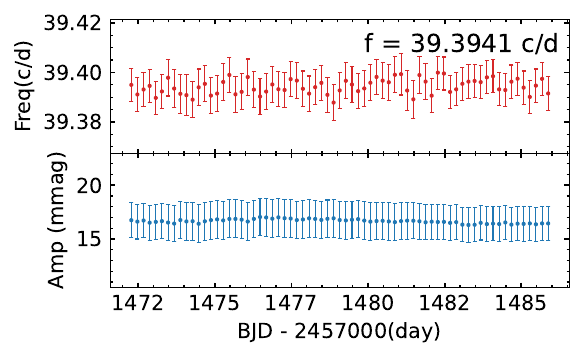} \\[-2pt]
\includegraphics[width=0.33\textwidth]{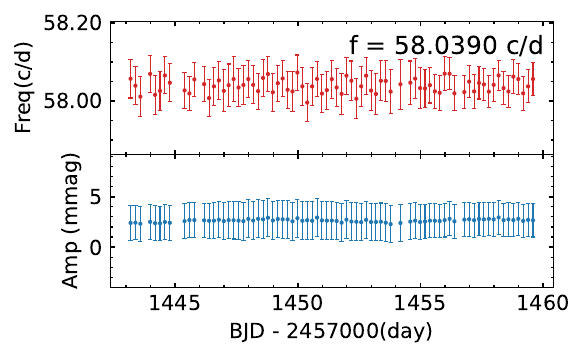} &
\includegraphics[width=0.33\textwidth]{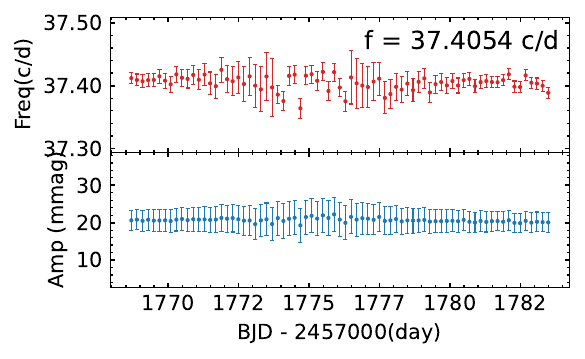} &
\includegraphics[width=0.33\textwidth]{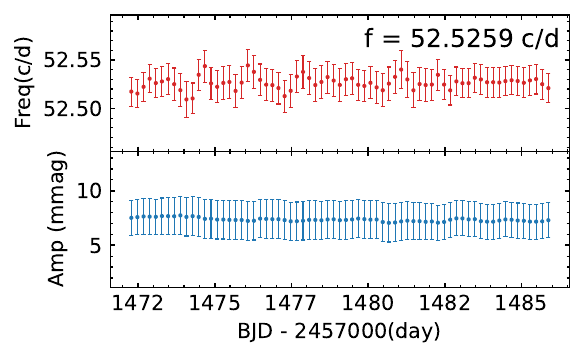} \\[-2pt]
\end{tabular}
\caption{Time-dependent amplitudes and instantaneous frequencies of the dominating frequency and harmonics for the three HADS targets, derived from a sliding-window analysis with an 8-day window and 0.2-day step.
\textit{Left column:} TIC~408074920;
\textit{middle column:} TIC~189714989;
\textit{right column:} TIC~34137913.
Each panel shows the temporal evolution of the amplitude or instantaneous frequency for the the dominating frequency and low-order harmonic modes.}
\label{fig:amp_time_all}
\end{figure*}

Amplitude variability analysis in \dsct stars has been extensively studied over the past decades (e.g. \citealt{Breger2006, Breger2008, Bowman2014, Bowman2016}). To investigate the time-dependent behaviour of the pulsation modes identified in Section~4, we performed a sliding Fourier transform (SFT) analysis using a fixed-frequency approach. The frequencies obtained from the global analysis were adopted and kept fixed, while possible variations in amplitude and phase were tracked as a function of time.

The light curves were divided into overlapping time segments with a duration of 8\,days and a step size of 0.2\,days. This choice represents a compromise between temporal resolution and the need for stable amplitude and phase estimates, and ensures that each segment contains a sufficient number of pulsation cycles of the dominant modes \citep{Niu2022}. Within each segment, a sinusoidal model with fixed frequency but free amplitude and phase was fitted using a non-linear least-squares method.

The uncertainties of the fitted amplitudes and phases were estimated from the residuals of the fit. For a sinusoidal signal in the presence of Gaussian noise, the amplitude uncertainty is given by $\sigma_A = \sqrt{2/N}\,\sigma_m$, where $N$ is the number of data points in the segment and $\sigma_m$ is the standard deviation of the residuals \citep{Montgomery1999}. Although the frequency was held fixed during the fitting, an effective frequency uncertainty was estimated from the phase variations as $\sigma_f = \sqrt{3}\,\sigma_A/(\pi A T_{\rm seg})$, where $A$ is the fitted amplitude and $T_{\rm seg}$ is the segment length.

Assuming that the phase variations are continuous and that any modulation occurs on timescales much longer than the pulsation period, the instantaneous frequency of each mode was derived from the time derivative of the phase according to $f_{\rm local} = f_{\rm global} + (1/2\pi)\,({\rm d}\phi/{\rm d}t)$. It should be noted that these instantaneous frequencies are not independently fitted parameters, but are derived quantities based on the phase evolution under the fixed-frequency assumption. To ensure the reliability of the derived instantaneous frequencies, we applied a filtering criterion and excluded points with frequency deviations exceeding 0.05 c/d from the global frequency, which is comparable to the Rayleigh frequency resolution of the dataset. The corresponding amplitude measurements and uncertainties were removed accordingly. The resulting amplitudes and instantaneous frequencies were then plotted as a function of time for each mode, producing the tracking diagrams shown in Figure~\ref{fig:amp_time_all}. For all three stars, the amplitudes and phases of the dominant modes remain largely stable within the observational uncertainties over the \textit{TESS} observing baseline. No evidence is found for strong or rapid amplitude or frequency modulation on timescales shorter than the data span. For the HADS stars in our sample, this result places quantitative upper limits on possible short-term variability and supports the interpretation that any detected side peaks in the frequency spectra are associated with long-term or weak modulation effects rather than pronounced time-dependent changes within a single \textit{TESS} sector.

\section{Asteroseismic Modeling}
\label{sec:astero}

\begin{figure*}[h]
\begin{center}
  \includegraphics[width=0.95\textwidth]{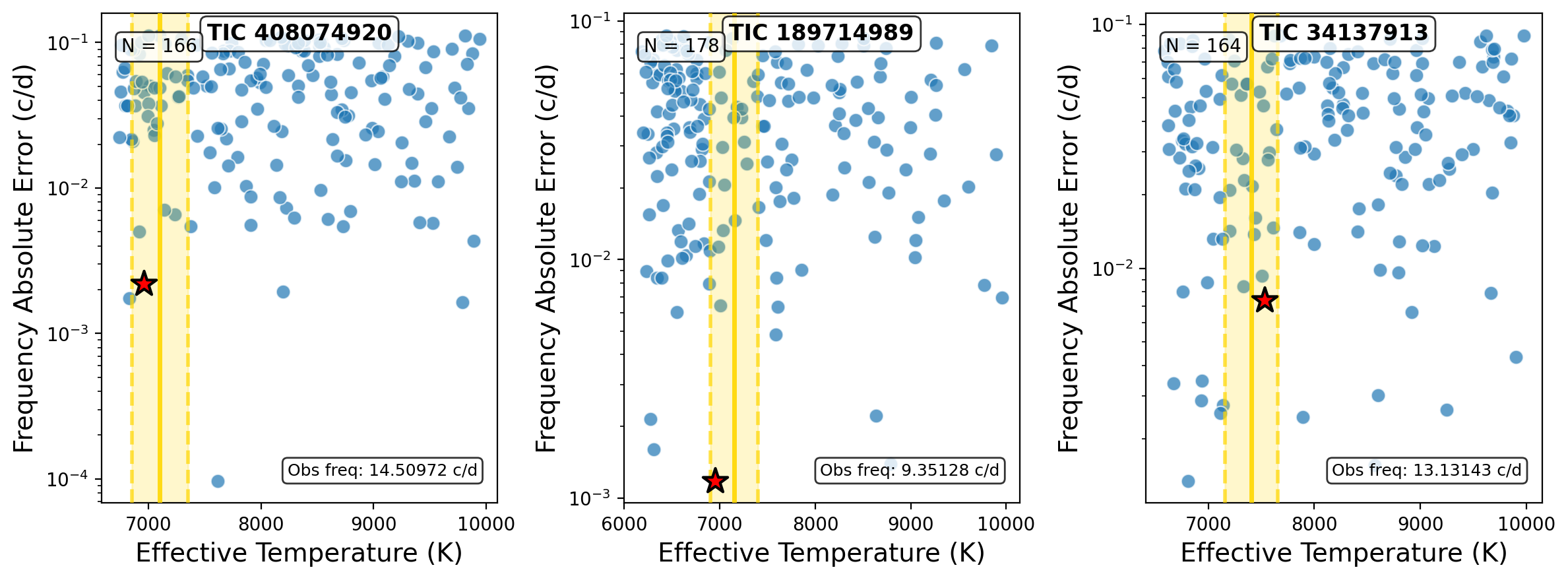}
  \caption{Distribution of the frequency difference between the modeled
and observed radial-mode frequencies as a function of effective temperature. Each panel shows one target star. Blue dots denote models with frequency differences smaller than 0.04 c/d, and the red star marks the best-matching model for each star. The yellow shaded region in each panel represents the $\pm 250 {\rm K}$ uncertainty range in effective temperature.}
    \label{fig:teff_error}
\end{center}
\end{figure*}

Stellar evolutionary and seismic models were constructed for three individual target stars using the one-dimensional stellar evolution code \texttt{MESA} (version 24.08.1; \citealt{Paxton2011,Paxton2013,Paxton2015,Paxton2018,Paxton2019,Jermyn2023}). Each star was modeled independently, and no assumption of a common set of stellar parameters was imposed.

All calculations were initialized from the pre-main-sequence phase, with the stellar age set to zero and the initial structures relaxed over 100 pre-main-sequence steps. The evolutionary sequences were terminated at central hydrogen exhaustion, defined by a central hydrogen mass fraction of $X_{\rm c} = 10^{-5}$. Nuclear energy generation was computed using the \texttt{o18\_and\_ne22.net} reaction network. Radiative opacities were adopted from the solar abundance mixture of \citet{Asplund2009}, employing the corresponding A09 opacity tables together with the low-temperature opacities \texttt{lowT\_fa05\_a09p}. The equation of state was taken from the default \texttt{MESA} configuration. Atmospheric boundary conditions were described using a $T$--$\tau$ relation based on the Eddington approximation with a varying-opacity atmosphere. Convection was treated using the standard mixing-length theory with the ML1 formulation, adopting a fixed mixing-length parameter of $\alpha_{\rm MLT} = 1.9$ \citep{Paxton2013}. Convective core overshooting was included using an exponential diffusive prescription applied at the upper boundary of the hydrogen-burning convective core, with an overshooting efficiency parameter of $f_{\rm ov} = 0.010$ \citep{Guo2019}. Effects due to rotation, magnetic fields, and microscopic diffusion were neglected.

For each target star, a preliminary grid of evolutionary models was computed to sample the parameter space around the observed stellar properties. The initial stellar mass was varied between $1.50$ and $2.50\,M_\odot$ in steps of $0.02\,M_\odot$, while the initial metallicity ranged from $Z = 0.001$ to $0.030$ with a step size of $0.002$. The initial helium abundance was assigned according to the relation $Y = 0.249 + 1.33Z$.

Adiabatic oscillation frequencies were calculated using the pulsation code \texttt{GYRE} \citep{Townsend2013}.
For each selected stellar model, only radial oscillation modes ($\ell = 0$) were considered. The frequencies were searched within the range 4-17 c/d.
To ensure that all eigenmodes within this range were identified, we used 200 trial frequencies as initial guesses for the root-finding procedure in \texttt{GYRE}.
The final eigenfrequencies are obtained through an iterative solution of the pulsation equations and are therefore not limited by the spacing of the initial trial frequencies.
The agreement between theoretical models and observations was evaluated by directly comparing the modeled fundamental radial-mode frequency with the observed frequency. Models were considered acceptable if the absolute frequency difference
was smaller than 0.04 c/d. This value is comparable to the Rayleigh frequency resolution of the TESS light curves used in this work. The distribution of the frequency differences as a function of effective temperature is shown in Figure~\ref{fig:teff_error}, where blue dots represent acceptable models and the red star marks the best-matching model for each target. The best-fit model parameters for TIC~408074920, TIC~189714989, and TIC~34137913 are summarized in Table~\ref{Tab:bestmodels}.

\begin{deluxetable}{ccccccccccc}
\renewcommand\arraystretch{1.2}
\tabletypesize{\small}
\setlength\tabcolsep{12pt}
\tablewidth{\textwidth}
\tablenum{6}
\tablecaption{Best-fit model parameters for TIC~408074920, TIC~189714989, and TIC~34137913. \label{Tab:bestmodels}}
\tablehead{
\colhead{TIC ID} &
\colhead{Mass $(M_\odot)$} &
\colhead{Z} &
\colhead{Age (yr)} &
\colhead{$T_{\mathrm{eff}}$ (K)} &
\colhead{$\log g$} &
\colhead{$R_\star (R_\odot)$} &
\colhead{$f_{\mathrm{obs}}$ (c/d)} &
\colhead{$f_{\mathrm{mod}}$ (c/d)}
}
\startdata
408074920 & 1.54 & 0.015 & 1.46$\times 10^9$ & 6955 &4.083&1.868& 14.50972 & 14.51192 \\
189714989 & 1.76 & 0.013 & 1.31$\times 10^9$ & 6947 & 3.839&2.645&9.35128 & 9.35009 \\
34137913  & 1.84 & 0.023 & 8.54$\times 10^8$ & 7532 &4.035&2.158& 13.13143 & 13.12404 \\
\enddata
\tablecomments{Best-fit stellar models within the observed $T_{\mathrm{eff}}$ range. $f_{\mathrm{obs}}$ and $f_{\mathrm{mod}}$ represent observed and model frequencies, respectively.}
\end{deluxetable}

\section{Discussion}
\label{sec:discussion}

\subsection{SED Constraints and Internal Consistency}

For all three targets, the SED fitting constrained by \emph{Gaia} parallaxes yields well-determined posterior distributions for the $T_{\rm eff}$, $\log g$, $A_V$, stellar radius, and distance. The absolute magnitudes inferred from the SED-based distances also agree with the empirical period–luminosity relation (Figure~\ref{fig:PL}) within the observed dispersion, reinforcing the interpretation that the dominant detected frequencies correspond to the fundamental radial mode. The placement of the stars in the H-R diagram (Figure~\ref{fig:HRD}), based on the combination of $T_{\rm eff}$ from the SED fits and luminosities from \citet{Stassun2019}, falls securely within the classical instability strip, consistent with their classification as $\delta$~Scuti stars \citep{Breger2000}.

In addition, the Fourier spectra reveal a well-developed harmonic structure extending to multiple orders, which is consistent with the strongly non-sinusoidal light-curve morphology and the large peak-to-peak amplitudes. These features are in agreement with the radii and effective temperatures derived from the SED fits and are commonly observed in classical HADS stars. Taken together, these independent observational characteristics indicate that the derived stellar parameters are against plausible systematic effects, such as extinction corrections, photometric zero-point offsets, or uncertainties in the adopted model grids, and provide a self-consistent description of the pulsational properties of the targets.

\subsection{Amplitude Variability and Possible Physical Drivers}
\label{sec:ampvar_discuss}

We performed a time-dependent analysis of the amplitude and phase of the fundamental frequency and its harmonic series using a fixed-frequency sliding window method.
For TIC~408074920, the amplitude of the first overtone mode shows a gradual decrease of approximately $\sim$4~mmag over the $\sim$27-day TESS baseline, while the frequency remains essentially stable. In contrast, TIC~34137913 shows no clear or systematic variations in either amplitude or frequency within the same time span.

Several physical mechanisms may contribute to the observed behavior. Unresolved beating with a nearby mode could mimic smooth amplitude changes \citep{Bowman2017}. However, the global Fourier spectra of TIC~408074920 show no convincing close companion to $f_0$ at the Rayleigh resolution ($\sim$0.036–0.037~c/d), favoring genuine damping or slow modulation. A similar phenomenon was reported for the $\delta$ Scuti star KIC 7106205, where the dominant mode exhibited an amplitude decrease of several mmag over the Kepler time span and was interpreted as a possible consequence of beating between unresolved close-frequency modes \citep{Bowman2014,Bowman2017}. The $\sim$4~mmag amplitude decrease observed in TIC~408074920 over a $\sim$20~day interval is qualitatively reminiscent of that behavior, although the relatively short TESS time baseline limits the ability to resolve extremely close frequency components. Nonlinear energy exchange among pulsation modes and their harmonics is also plausible \citep{Dziembowski1985,Breger2014,Bowman2014,Bowman2015,Niu2024}, as TIC~408074920 exhibits harmonics up to at least $5f_0$, and anti-correlated amplitude variations between the parent mode and harmonics could occur on dynamical timescales.

This mechanism resonates with the recent discovery of nonlinear interactions in another HADS star, KIC 6382916, which challenges the stability of independent pulsation modes \citep{Niu2024}. That study found that amplitudes and frequencies of pulsation modes and their harmonics can undergo drastic changes over approximately 20 days, revealing the complexity of short-timescale mode interactions.

\begin{figure*}[htbp!]
\begin{center}
  \includegraphics[width=0.85\textwidth]{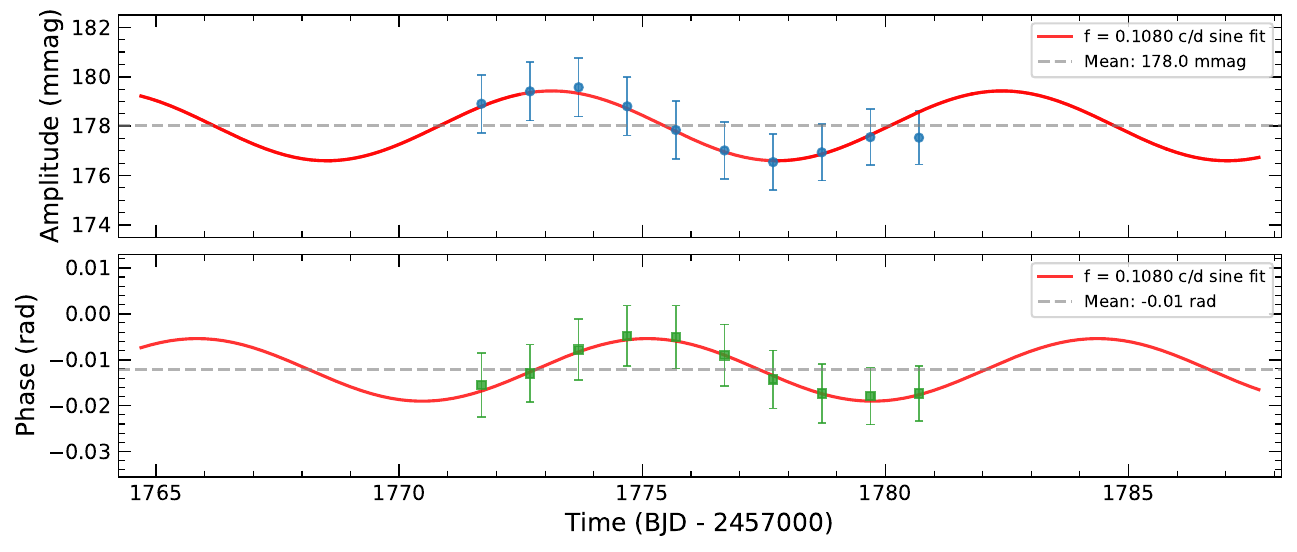}
  \caption{Time-dependent amplitudes and phases of the fundamental for TIC~189714989, derived from a sliding-window analysis with an 14-day window and 1-day step.}
    \label{fig:1898modu}
\end{center}
\end{figure*}

Geometric effects, such as oblique pulsation or rotational modulation, can produce side peaks and induce gradual variations in amplitude and phase \citep{Breger2011, Yang2018}. In the amplitude spectrum of TIC~189714989, symmetric sidebands separated by $f_m\simeq0.1197$~c/d are detected, consistent with amplitude or phase variability. However, when applying a SFT with an 8-day window, which is comparable to the estimated modulation period of $P_m\sim8.35$ days, the corresponding sideband structure could not be clearly resolved in the time-frequency spectra. Notably, by refining the analysis through the use of a 14-day time segment and a 1-day sliding step, we successfully resolved significant periodic modulations in both amplitude and phase for TIC~189714989 (see Figure~\ref{fig:1898modu}). The derived modulation frequency of 0.1080 c/d aligns closely with both the observed frequency splitting and the signal at $f_{27}$, within the frequency resolution limits.

The presence of symmetric sidebands together with the co-modulation of amplitude and phase suggests a modulation behavior reminiscent of the Blazhko effect commonly observed in RR Lyrae stars \citep{Kolenberg2011,Guggenberger2011}. However, given the limited $\sim$27-day TESS baseline, the physical origin of this modulation cannot be firmly established. Alternative explanations, such as binarity or geometric effects, may also account for the observed variability. Nevertheless, a binary interpretation appears unlikely. The Gaia DR3 solution for TIC~189714989 yields a renormalized unit weight error (RUWE) of 1.143 \citep{Gaia2023}, which indicates no significant astrometric excess noise and does not support the presence of an unresolved binary companion. In addition, the observed modulation period of $\sim$8.3 days would imply a very short orbital period if interpreted as binary-induced variability, which would typically produce detectable signatures such as ellipsoidal variations or additional photometric modulation that are not evident in the current TESS light curve.

\subsection{Placement in the period–luminosity diagram}

\begin{figure}[htbp!]
\begin{center}
  \includegraphics[width=0.4\textwidth]{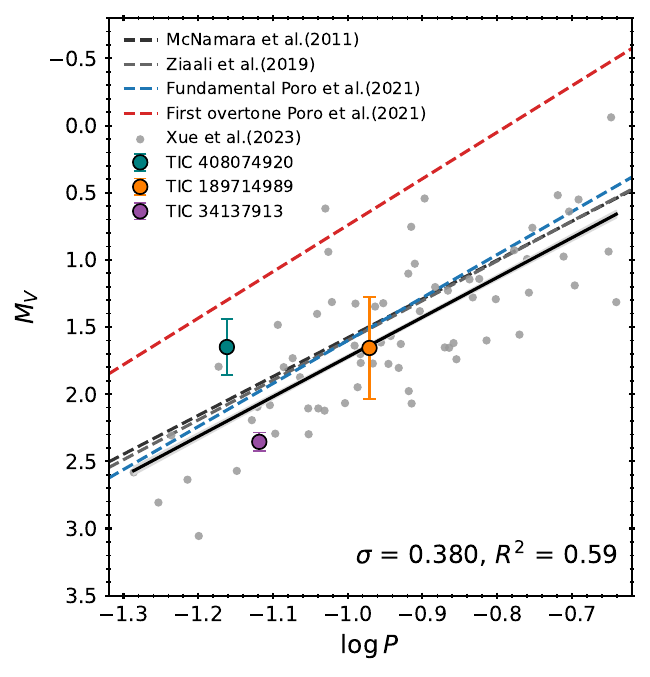}
  \caption{Period–luminosity relation for HADS stars from \citep{Xue2023}. Gray points show the full sample; the red line is the median MCMC linear fit, with the shaded area indicating the 16–84th percentile uncertainties. Highlighted stars TIC~408074920 (teal), TIC~189714989 (orange), and TIC 34137913~(purple) are marked, with only TIC~408074920 is indicated as F1, while the others are F0. The black dashed line is the period–luminosity relation for high-amplitude \dsct stars derived by \citet{2011AJ....142..110M}, while the period–luminosity relation obtained by \citet{Ziaali2019} is shown by the gray dashed line. The remaining dashed lines represent different modes of the period–luminosity relation, as derived by \citet{2021PASP..133h4201P}.}
    \label{fig:PL}
\end{center}
\end{figure}

Figure~\ref{fig:PL} compares our three targets with the empirical period–luminosity relation of HADS compiled by previous works (gray points and regression line; see caption).
All three stars lie within the intrinsic dispersion of the relation (the fit in Figure~\ref{fig:PL} has a scatter of $\sigma\simeq0.38$\,mag), and their residuals show no systematic offset toward either brighter or fainter magnitudes.
This supports the identification of the dominant peaks as fundamental radial modes (Sections~\ref{subsec:tic408074920}--\ref{subsec:tic34137913}) and implies that, despite their different frequency spectra, these stars conform to the same population used to calibrate HADS as distance indicators.
The consistency across targets also suggests that short-term amplitude variability (Section~\ref{sec:ampvar_discuss}) does not strongly bias period-based luminosity estimates on the timescale covered by \emph{TESS} sectors.

\subsection{Consistency Between SED and Seismic Stellar Parameters}

\begin{figure}[htbp!]
\begin{center}
  \includegraphics[width=0.45\textwidth]{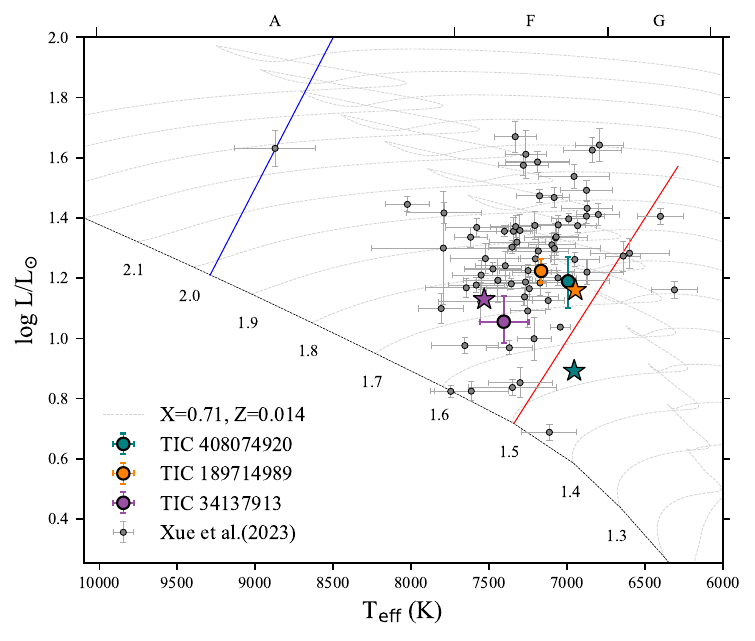}
  \caption{H-R diagram of the newly discovered HADS stars. Background HADS stars from \citep{Xue2023} are plotted in gray, with our three targets TIC~408074920 (teal), TIC~189714989 (orange), and TIC~34137913 (purple)—highlighted with error bars. Effective temperatures are from our SED fitting, and luminosities from \citet{Stassun2019}. Star symbols indicate the positions of the best-fitting MESA models. The observational instability strip is from \citet{Murphy2019}. The evolutionary tracks were calculated in MESA v10108 with X = 0.71 and Z = 0.01.}
    \label{fig:HRD}
\end{center}
\end{figure}

The fundamental properties of TIC~408074920, TIC~189714989, and TIC~34137913 were investigated using two complementary approaches: broadband SED fitting combined with \textit{Gaia}DR3 parallaxes, and stellar evolutionary and seismic modelling with \textsc{MESA} and \textsc{GYRE}. A comparison of the results from these largely independent methods provides a critical consistency check on the derived stellar parameters. In Figure~\ref{fig:HRD}, the three targets are placed on the HR diagram using the effective temperatures inferred from SED fitting and luminosities adopted from the literature. The coloured circles mark these values, while the coloured stars correspond to the best-fitting seismic models.

The effective temperatures obtained from the seismic best-fit models are broadly consistent with the SED-based estimates, with differences limited to a few hundred Kelvin. Consequently, the circle–star pairs for each target show only modest horizontal offsets in the HR diagram. Such offsets are expected given the different sensitivities of the two methods, as SED temperatures are constrained by the global spectral energy distribution, while evolutionary models yield temperatures that are self-consistent with the stellar structures required to reproduce the observed pulsation frequencies \citep{Chen2025}.

For TIC~189714989 and TIC~34137913, the stellar parameters derived from the seismic models are broadly consistent with the SED-based results, with the corresponding circle and star symbols occupying similar regions of the HR diagram. This agreement indicates that the combination of \textit{Gaia} parallaxes, bolometric fluxes, and the fundamental radial-mode frequencies yields a coherent solution for the stellar luminosity and mean density.

In contrast, TIC~408074920 shows a significant offset between the two determinations in the HR diagram, particularly in luminosity. The seismic model places the star at a substantially lower luminosity than the value inferred from the SED analysis, producing a noticeable vertical separation between the circle and star symbols. This discrepancy is consistent with the difference found in the stellar radius estimates, with the seismic radius ($\approx 1.87~R_\odot$) being significantly smaller than the SED-derived value ($\approx 2.64~R_\odot$). Because luminosity scales approximately as $L\propto R^{2}T_{\rm eff}^{4}$, the smaller seismic radius naturally leads to a lower model luminosity. The discrepancy likely reflects the distinct information content of the two methods: the SED radius depends primarily on flux, extinction, and distance, whereas the seismic radius is tightly constrained by the mean stellar density implied by the fundamental radial-mode frequency. Additional systematics, such as uncertainties in extinction, unresolved binarity, or rotational effects on pulsation frequencies, may also contribute to the radius discrepancy \citep{Huber2017a}.

The surface gravities derived from seismic modelling differ modestly from those obtained via SED fitting, which is expected given the weak sensitivity of broadband photometry to $\log g$ \citep{Vines2022}. Similarly, metallicity is poorly constrained by SED fitting, and the $Z$ values inferred from the seismic models should be regarded as model-dependent parameters rather than precise abundance measurements. Robust constraints on the chemical composition will require high-resolution spectroscopy.

The stellar masses and ages derived in this work are exclusively constrained by the evolutionary and seismic modelling, placing all three targets within the intermediate-mass regime characteristic of $\delta$Scuti stars and indicating that they are on or slightly evolved from the main sequence. Overall, the comparison between the SED-derived and seismic parameters indicates a broadly consistent picture for two of the targets, while TIC~408074920 remains the only case where the luminosity inferred from the two approaches differs substantially. The combined use of SED fitting and asteroseismic modelling can therefore contribute to the characterization of the global and internal properties of $\delta$~Scuti stars.

\section{Summary}
\label{sec:summary}

In this work we carried out a detailed photometric analysis of three HADS stars observed by the \textit{TESS} mission: TIC~408074920, TIC~189714989, and TIC~34137913. The high precision and continuous coverage of the \textit{TESS} data allow us to resolve the detailed pulsation structures of these large-amplitude radial pulsators and to investigate the complexity of their oscillation behaviour.

For TIC~408074920, the pulsation spectrum is dominated by a single frequency at $f_1 = 14.509721$~c/d together with a clear harmonic sequence extending to $5f_0$. The derived pulsation constant is more consistent with the identification of this mode as the first overtone radial oscillation, confirming the star as a classical HADS pulsator. In addition, a gradual decrease in pulsation amplitude is detected over the \textit{TESS} observing window, with the peak-to-peak amplitude decreasing by about 4\,mmag within approximately 20 days. This behaviour indicates the presence of short-term amplitude variability in a star otherwise dominated by a stable radial pulsation mode.

TIC~189714989 exhibits a significantly more complex frequency spectrum. Besides the fundamental radial mode ($f_1 = 9.35128$~c/d) and its harmonics up to 8F0, nearly symmetric side peaks are detected around the dominant mode and its harmonics with a separation of $f_{\rm m} \simeq 0.1197$~c/d. This pattern clearly indicates amplitude and phase modulation of the dominant pulsation. After examining possible explanations, the binary scenario can be excluded, implying that the modulation is intrinsic to the pulsation itself. In addition, an independent pulsation mode with a period ratio $P_{24}/P_0 \approx 0.662$ is detected. This ratio is inconsistent with radial overtone modes and therefore suggests the presence of a low-degree non-radial p-mode. The coexistence of radial pulsation, modulation side peaks, and a non-radial mode is consistent with previous studies showing that HADS stars can exhibit more complex pulsation behaviour than traditionally assumed.

For TIC~34137913, the pulsation spectrum is also dominated by the fundamental radial mode at $f_1 = 13.131429$~c/d together with harmonics up to 6F0. An additional independent mode at $f_7 = 17.6174$~c/d is detected with a period ratio $P_7/P_0 \sim 0.746$. This value lies slightly below the canonical ratio expected for the first radial overtone, suggesting that the additional signal may correspond to a low-degree non-radial p-mode or a higher-order radial oscillation. The combination frequencies involving these modes are also identified, which are commonly interpreted as signatures of nonlinear interactions between pulsation modes.

The spectral energy distribution modelling provides independent estimates of the effective temperature, stellar radius, and extinction for the three targets. The derived parameters place all stars within the classical instability strip and confirm their classification as $\delta$~Scuti variables. Taken together, these results provide new observational constraints on the pulsation behaviour of HADS stars based on high-precision \textit{TESS} photometry. Although the pulsation spectra of all three stars are dominated by the fundamental radial mode or first overtone, the data also exhibit additional phenomena that extend beyond the classical picture of HADS stars as strictly stable radial pulsators, consistent with previous studies that have reported similar complexity.

In particular, we detect short-term amplitude variability in TIC~408074920, intrinsic amplitude and phase modulation of the dominant radial mode in TIC~189714989, and additional independent pulsation modes in the three stars. While the presence of non-radial modes in HADS stars has been reported in previous studies, the \textit{TESS} data analysed here clearly show that large-amplitude radial pulsation can coexist with modulation phenomena and additional pulsation modes in the same stars. These results highlight the complex dynamical behaviour that can occur even in classical HADS pulsators and provide new observational constraints for theoretical models of mode selection and nonlinear pulsation in $\delta$~Scuti stars. Finally, the results here reported emphasize the importance of ultra-precise data and long-term monitoring to study the HADS stars. In this sense, the PLATO mission \citep{PLATO} will provide a unique opportunity to shed some light into the nature of this variability. However, innovative approaches such as the application of complex network analysis \citep{Ziaali25} might be necessary to explore the underlying physical mechanisms and characterise properly the dynamics here shown.

\begin{acknowledgments}

We are grateful to the anonymous reviewer for the professional and thorough suggestions. Guojie Feng is supported by the Chinese Academy of Sciences (CAS) "Light of West China" Program (No. 2022-XBQNXZ-016). Chenglong Lv acknowledges support from the National Natural Science Foundation of China (NSFC; No. 1250030315), the Natural Science Foundation of Xinjiang Uygur Autonomous Region (No. 2025D01B176). Ali Esamdin is supported by the National Key R\&D program of China for the Intergovernmental Scientific and Technological Innovation Cooperation Project under No. 2022YFE0126200, and the Tian-shan Talent Training Program (No. 2023TSYCLJ0053). This work is also supported by Xinjiang Uygur Autonomous Region Introduced Project "Tianchi talent". JPG acknowledges financial support from project PID2023-149439NB-C42 from the ‘Proyectos de Generación de Conocimiento’ and from the Severo Ochoa grant CEX2021-001131-S funded by MICIU/AEI/10.13039/501100011033 and FEDER, EU.

This paper includes data collected with the \textit{TESS} mission, obtained from the Mikulski Archive for Space Telescopes (MAST) at the Space Telescope Science Institute (STScI). Funding for the \textit{TESS} mission is provided by the NASA Explorer Program. STScI is operated by the Association of Universities for Research in Astronomy, Inc., under NASA contract NAS~5-26555. We acknowledge the use of public \textit{TESS} data from MAST: \url{https://archive.stsci.edu/}. This work also presents results from the European Space Agency (ESA) space mission Gaia. Gaia data are being processed by the Gaia Data Processing and Analysis Consortium (DPAC). Funding for the DPAC is provided by national institutions, in particular the institutions participating in the Gaia MultiLateral Agreement (MLA). The Gaia mission website is https://www.cosmos.esa.int/gaia. The Gaia archive website is https://archives.esac.esa.int/gaia.

\software{
\texttt{astropy} \citep{Astropy2013,Astropy2018,Astropy2022},
\texttt{matplotlib} \citep{Hunter2007},
\texttt{numpy} \citep{Harris2020},
\texttt{scipy} \citep{Virtanen2020}
\texttt{Lightkurve} \citep{Lightkurve2018}
\texttt{ASTROARIADNE} \citep{Vines2022},
\texttt{FELIX} \citep{Charpinet2010,Zong2016},
}

\end{acknowledgments}

\appendix
\section{Significance Criterion for Frequency Detection}

\begin{figure}[htp!]
\begin{center}
  \includegraphics[width=0.55\textwidth]{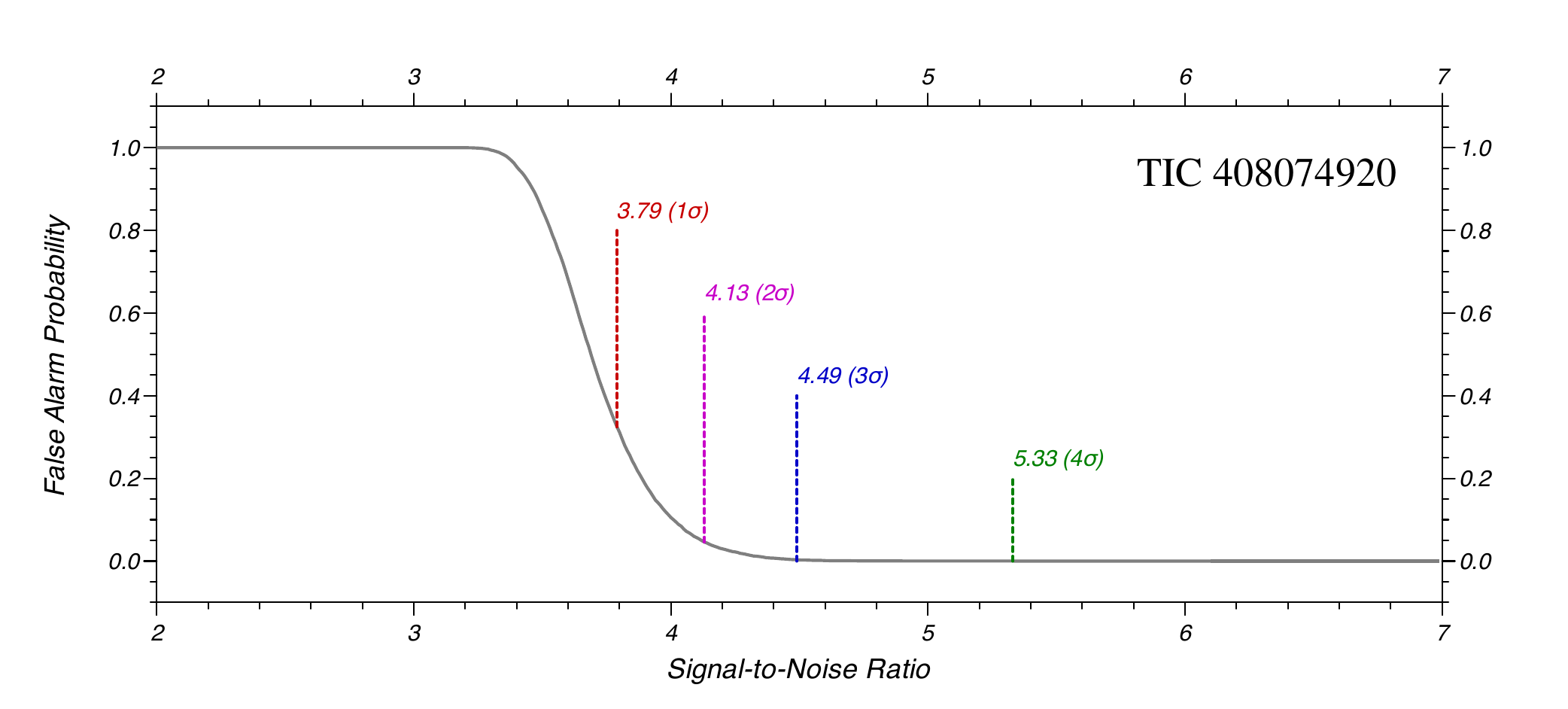}
  \includegraphics[width=0.55\textwidth]{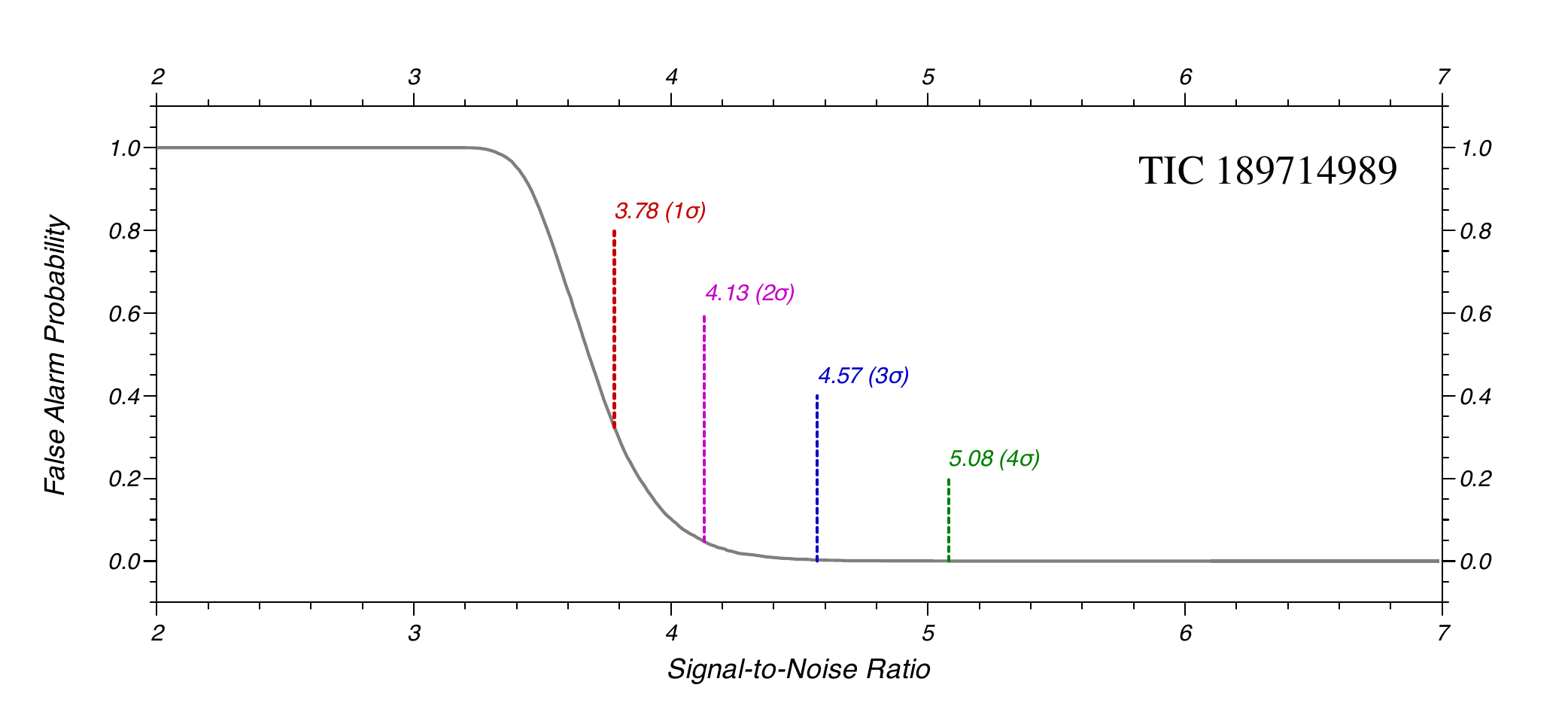}
  \includegraphics[width=0.55\textwidth]{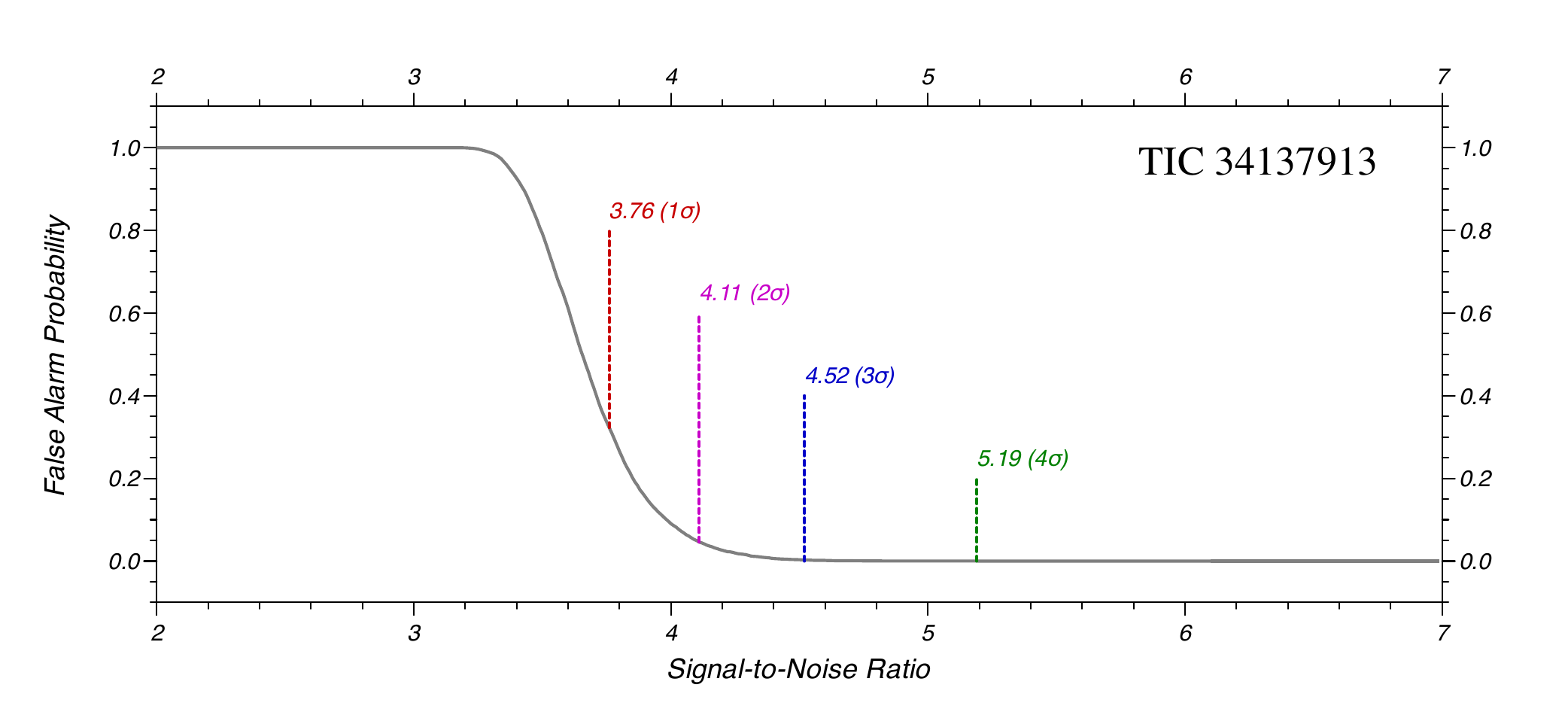}
  \caption{FAP curves for the three HADS stars analyzed in this work.}
    \label{fap}
\end{center}
\end{figure}

For each target, the false alarm probability (FAP) was evaluated independently, and the corresponding signal-to-noise ratio (S/N) at the $4\sigma$ significance level was determined. For all three HADS stars, the $4\sigma$ thresholds correspond to S/N values slightly above 5.0. Based on this result, we adopted a uniform and conservative criterion of $\mathrm{S/N} > 5.2$ for the identification of statistically significant frequencies throughout this work (Figure~\ref{fap}).

\clearpage

\bibliography{references}{}
\bibliographystyle{aasjournalv7}

\end{document}